\documentclass[twocolumn]{aastex61}
\usepackage{amsmath}
\usepackage{CJK}

\newcommand{\dif}{{\rm d}}

\newcommand{\yr}{{\rm year}}
\newcommand{\sigmai}{{\sigma_{i,5}}}

\shorttitle{Intrinsic Architecture of Planetary Systems}
\shortauthors{Zhu et al.}

\begin{document}
\begin{CJK*}{UTF8}{gbsn}

\title{About 30\% of Sun-like Stars Have \emph{Kepler}-like Planetary Systems: A Study of their Intrinsic Architecture}

\author{Wei~Zhu (祝伟)}
\affil{Canadian Institute for Theoretical Astrophysics, University of Toronto, 60 St. George Street, Toronto, ON M5S 3H8, Canada}
\correspondingauthor{Wei Zhu}
\email{weizhu@cita.utoronto.ca}

\author{Cristobal Petrovich}
\affil{Canadian Institute for Theoretical Astrophysics, University of Toronto, 60 St. George Street, Toronto, ON M5S 3H8, Canada}
\affil{Centre for Planetary Sciences, Department of Physical \& Environmental Sciences, University of Toronto at Scarborough, Toronto, Ontario M1C 1A4, Canada}

\author{Yanqin Wu (武延庆)}
\affil{Department of Astronomy and Astrophysics, University of Toronto, 50 St. George Street, Toronto, ON M5S 3H4, Canada}

\author{Subo Dong (东苏勃)}
\affil{Kavli Institute for Astronomy and Astrophysics, Peking University, Yi He Yuan Road 5, Hai Dian District, Beijing 100871, China}

\author{Jiwei Xie (谢基伟)}
\affil{School of Astronomy and Space Science \& Key Laboratory of Modern Astronomy and Astrophysics in Ministry of Education, Nanjing University, Nanjing 210093, China}

\begin{abstract}
    We constrain the intrinsic architecture of \emph{Kepler} planetary systems by modeling the observed multiplicities of the transiting planets (tranets) and their transit timing variations (TTVs). We robustly determine that the fraction of Sun-like stars with \emph{Kepler}-like planets, $\eta_{\rm Kepler}$, is $30\pm3\%$. Here \emph{Kepler}-like planets are planets that have radii $R_{\rm p} \gtrsim R_\oplus$ and orbital periods $P<400$~days. Our result thus significantly revises previous claims that more than 50\% of Sun-like stars have such planets.
    Combining with the average number of \emph{Kepler} planets per star ($\sim0.9$), we obtain that on average each planetary system has $3.0\pm0.3$ planets within 400 days.
    We also find that the dispersion in orbital inclinations of planets within a given planetary system, $\sigma_{i,k}$, is a steep function of its number of planets, $k$. This can be parameterized as $\sigma_{i,k}\propto k^\alpha$ and we find that $-4<\alpha<-2$ at 2-$\sigma$ level.
    Such a distribution well describes the observed multiplicities of both transits and TTVs with no excess of single-tranet systems. Therefore we do not find evidence supporting the so-called ``\emph{Kepler} dichotomy.''
    Together with a previous study on orbital eccentricities, we now have a consistent picture: the fewer planets in a system, the hotter it is dynamically.
    We discuss briefly possible scenarios that lead to such a trend. Despite our Solar system not belonging to the \emph{Kepler} club, it is interesting to notice that the Solar system also has three planets within 400 days and that the inclination dispersion is similar to \emph{Kepler} systems of the same multiplicity.
\end{abstract}                                     


\section{Introduction} \label{sec:introduction}

    The term ``planet occurrence rate'' has two different interpretations: the average number of planets per star, and the fraction of stars with planets. These two quantities are different, unless all planetary systems have only one planet. With transiting planets (tranets, \citealt{Tremaine:2012}) from surveys such as the \emph{Kepler} mission \citep{Borucki:2010}, one can constrain the first but not the second \citep{Youdin:2011}, unless assumptions of the intrinsic architecture (e.g., the orbital inclination distribution and/or the intrinsic multiplicity function) are made. This is because, to determine the average number of planets per star, one needs to compute the probability that individual planet transits, which only involves the orbital period of the planet (or more precisely, the ratio of stellar radius to orbital separation, $R_\star/a$). The distribution of orbital periods can be reconstructed from the orbital periods of observed tranets. However, to determine the fraction of stars with planets, one needs to compute the probability that a given star has at least one tranet. This involves the orbital inclinations of all planets around this star, in addition to the orbital periods of these planets. The distribution of planetary inclinations cannot be reconstructed from the tranet sample, because by definition all tranets have $\sim90^\circ$ inclinations.

The commonly accepted result says that more than $50\%$ of Sun-like stars have \emph{Kepler}-like planets \citep{Fressin:2013,Petigura:2013,Winn:2015}. However, they used the transit probability of the innermost planet (as in \citealt{Fressin:2013}, or the most easily detected one as in \citealt{Petigura:2013}) as the probability that at least one planet transits. Therefore, their estimates are only valid under the assumption that all detected planets are in multi-planetary systems on coplanar orbits.

Either using the combined constraints of \emph{Kepler} and Radial Velocity (RV) data \citep{Tremaine:2012, Figueira:2012} or transit duration distributions normalized by orbital velocities \citep{Fabrycky:2014, Fang:2012}, multi-tranet systems are found to be on average nearly co-planar. However, these methods cannot be applied to the {\it Kepler} single-tranet systems, which contribute over half of the detected tranets.

\citet{Lissauer:2011} modeled the observed multiplicity function of \emph{Kepler} and found the single-tranet systems are in excess to a single simulated underlying planet population. This is sometimes called the ``Kepler dichotomy'' \citep{Johansen:2012,Ballard:2016}. However, \citet{Tremaine:2012} showed that modeling the observed multiplicity function from \emph{Kepler} data alone cannot arrive at a reliable conclusion on the intrinsic multiplicity function due to its degeneracy with inclination distribution.

The transit timing variation (TTV) technique can help break the degeneracy between intrinsic multiplicity function and inclination distribution \citep[e.g.][]{Xie:2014}. Although TTV is the behavior of the transiting planet, it can reveal the existence of the non-transiting companion \citep{Holman:2005,Agol:2005}. If there is indeed a large population of intrinsic singles, then transiting planets in the single-tranet systems should have considerably smaller probability to show TTV signals than the transiting planets in multi-tranet systems. However, this is not supported by the large and uniform TTV catalogs. For example, \citet{Holczer:2016} found that of the total 260 \emph{Kepler} planets that showed TTV signals, 121 were in single-tranet systems. The larger TTV catalog of \citet{Ofir:2018} that is more sensitive to smaller TTV amplitudes gives a similar result. Both strongly indicate that transiting planets in transit singles and transit multiples have similar probability to show TTV signals, and therefore that there is no large population of intrinsic singles.

Another evidence against the assumption that all \emph{Kepler} planets are coplanar comes from the study of the distribution of planet eccentricities. Using the distribution of transit durations, \citet{Xie:2016} found that single tranets have on average substantially larger eccentricities than multiple tranets. Because the dispersions of orbital eccentricities and inclinations are generally expected to be correlated \citep[e.g.][]{Ida:1993}, this result suggests that systems with fewer number of planets may have larger mutual inclinations.

If a significant fraction of planets are in multi-planet systems with larger mutual inclinations than previously thought, then the probability that one star is seen to have at least one tranet increases. Therefore, the total fraction of Sun-like stars with at least one \emph{Kepler}-like planet will decrease.

For similar reasons, the statistical studies based on the radial velocity (RV) samples also overestimated the fraction of stars with planets, by using the probability to detect the most detectable planet for the probability to detect at least one planet \citep{Cumming:2008,Mayor:2011}. Although this overestimation is less severe for giant planets because of their low multiplicity rate, it can significantly reduce the fraction of stars with lower-mass planets (such as super-Earths).

In this study, we combine the information of transiting planets and their non-transiting companions as inferred by TTVs, to constrain the intrinsic architecture of planetary systems. We focus on Sun-like stars in this paper. In Section~\ref{sec:sample} we construct the transit and TTV multiplicity functions based on a homogeneous sample. We then forward model these functions to constrain the intrinsic architecture in Section~\ref{sec:model}. Our results are presented in Section~\ref{sec:results}, and discussed in more details in Section~\ref{sec:discussion}.

\section{\emph{Kepler}-LAMOST Sample} \label{sec:sample}

To select Sun-like (FGK-type dwarf) stars for our study, we rely on the spectroscopic data from the Large Sky Area Multi-Object Fiber Spectroscopic Telescope (LAMOST, also known as Goushoujing Telescope, \citealt{Cui:2012,Zhao:2012}), which had surveyed over 30\% of all \emph{Kepler} targets by 2017 (DR4), with no bias toward planet hosts \citep{Decat:2015,Ren:2016}. The derived stellar parameters are accurate at least for main-sequence stars, as previous studies \citep{Dong:2014,Xie:2016} have shown. The sample selection is similar to \citet{Xie:2016} and \citet{Dong:2017}. In short, we find 30,759 stars with effective temperature $T_{\rm eff}$ in the range $4700-6500$ K and stellar surface gravity $\log{g}>4.0$ (in cgs unit), based on the stellar parameters derived by the LAMOST official pipeline \citep[LASP,][]{Luo:2015,Xie:2016}. We then cross-match this stellar catalog with the planet candidate table from \emph{Kepler} data release 23 \citep{Mullally:2015}, and find 1635 KOIs. Then we remove KOIs that meet the following criteria:
\begin{enumerate}
    \item Identified by \citet{Mullally:2015}, \citet{Coughlin:2016}, and \citet{Thompson:2017} as false positives; 484 are removed.
    \item KOIs with transit S/N$<7.1$ according to \citet{Mullally:2015}; 108 are removed.
    \item KOIs with $P>400~$days; 34 are removed.
    \item KOIs with $R_p>20~R_\oplus$; here $R_p$ is computed using LAMOST stellar parameters. 108 are removed.
    \item KOIs that are in single-tranet systems and have large False Positive Probabilities (FPP $>68\%$, \citealt{Morton:2016}); 74 are removed.
\end{enumerate}
The last criterion is not applied to the multi-tranet systems, which overall have very low false positive rates \citet{Lissauer:2012}. As \citet{Morton:2016} pointed out, their FPPs for multi-tranet systems could have been inflated by the effect of unidentified TTVs. Indeed, Kepler-23b \citep{Ford:2012} and Kepler-50b \citep{Steffen:2013} are both confirmed planets, but have FPP$>0.68$ according to \citet{Morton:2016}. This criterion is applied to the single-tranet systems, because of their overall high false positive rates and low TTV fractions. Of the 74 single-tranets removed, only two have TTV signals according to \citet{Holczer:2016}, the inclusion or exclusion of which does not affect our results.

\begin{figure}
\centering
\epsscale{1.2}
\plotone{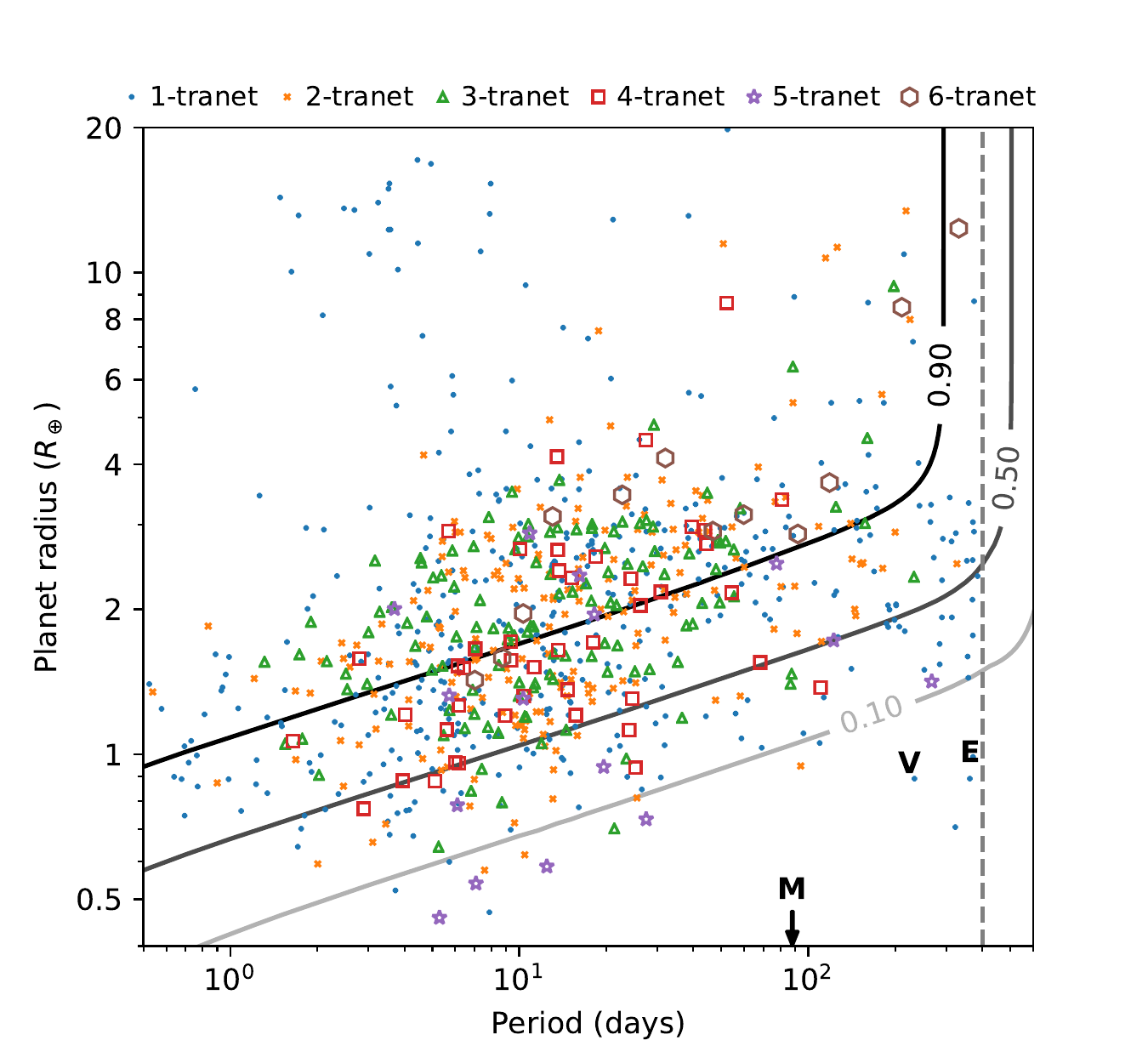}
    \caption{This plot demonstrates the radii and orbital periods of transiting planets (tranets) in our \emph{Kepler}-LAMOST sample. Tranets in different multiples are shown with different symbols and colors. We also over-plot the average efficiency of the \emph{Kepler} detection pipeline \citep{Burke:2015} as well as the positions of Solar system planets (Mercury, Venus, and Earth). The vertical dashed line indicates the period boundary (400 days).
\label{fig:sample}}
\end{figure}

\begin{figure}
\centering
\epsscale{1.2}
\plotone{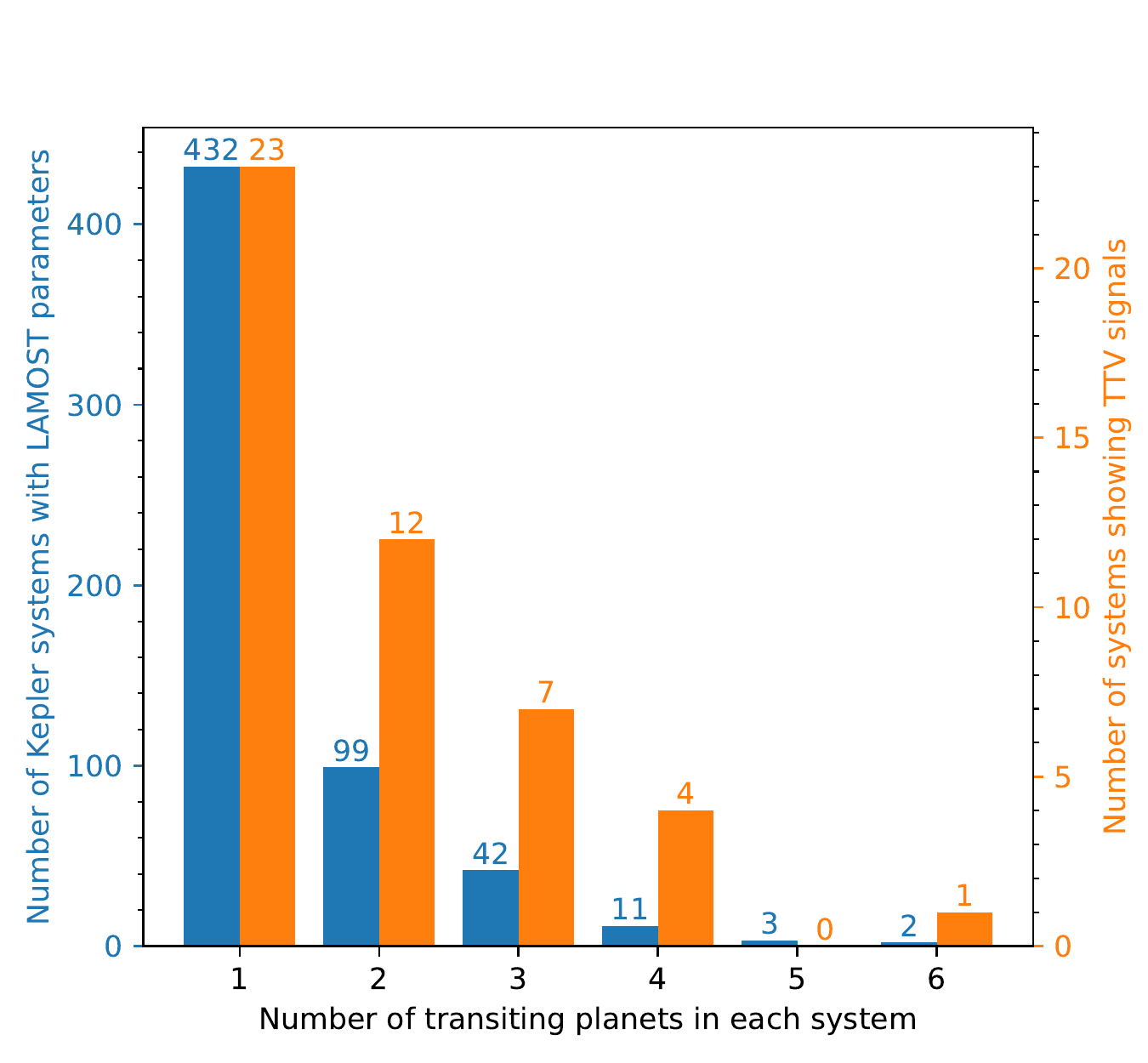}
\caption{Transit and TTV multiplicity functions constructed based on our sample. Here the TTV multiplicity means the number of systems with $j$ transiting planets and at least one of them showing TTV signal.
\label{fig:multiplicities}}
\end{figure}

In the end, we have 827 planets (or planet candidates) around 589 stars. The transit multiplicity function, denoting the number of systems as a function of number of tranets in each system, is $(N_1,~N_2,~N_3,~N_4~,N_5,~N_6)=(432,~99,~42,~11,~3,~2)$, and no system with more than six tranets. We show in Figure~\ref{fig:sample} the radii and orbital periods for tranets in our sample. The observed transit multiplicity function is illustrated in Figure~\ref{fig:multiplicities}.

\begin{figure*}[t!]
\centering
\epsscale{1.2}
\plotone{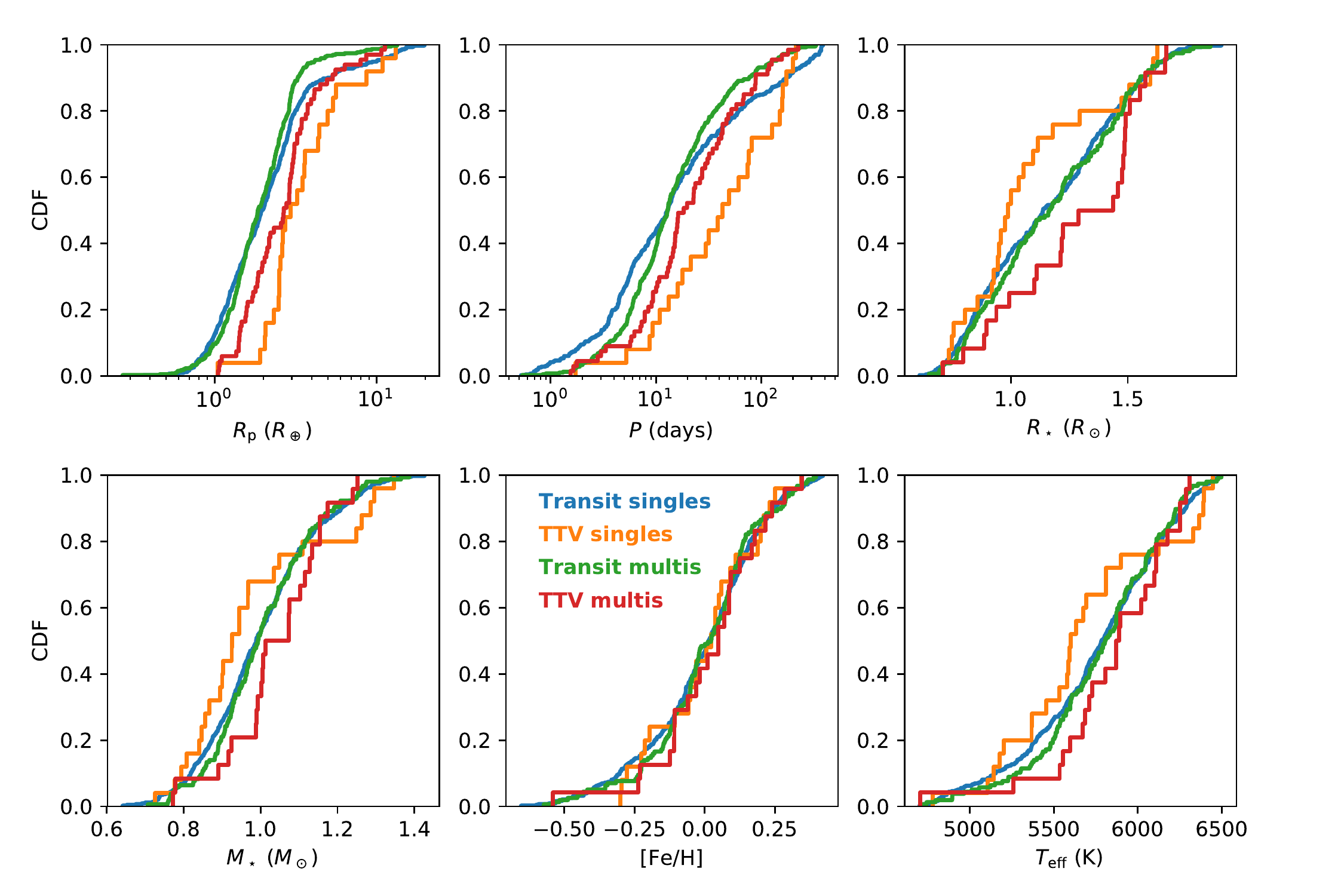}
\caption{Cumulative distributions of planetary and stellar parameters for four subsamples of planets in the \emph{Kepler}-LAMOST sample.
\label{fig:comparison}}
\end{figure*}

\begin{deluxetable*}{lcccccc}
\tablecaption{The two-sample KS test $p$ values for different combinations of subsamples and different (planetary and stellar) parameters.
\label{tab:p-values}}
\tablehead{
\colhead{Input subsamples} & \colhead{$R_{\rm p}$} & \colhead{$P$} & \colhead{$R_\star$} & \colhead{$M_\star$} & \colhead{[Fe/H]} & \colhead{$T_{\rm eff}$}}
\startdata
Transit singles \& Transit multis & 0.055 & 0.003$^a$ & 0.80 & 0.52 & 0.66 & 0.55 \\
Transit singles \& TTV singles & \nodata & \nodata & 0.09 & 0.15 & 0.96 & 0.15 \\
Transit multis \& TTV multis & \nodata & \nodata & 0.10 & 0.08 & 0.99 & 0.63 \\
TTV singles \& TTV multis & 0.08 & 0.09 & 0.014$^b$ & 0.005$^b$ & 0.95 & 0.074 \\
\enddata
\tablecomments{$^a$ This small $p$ value is likely due to the geometric effect. See Appendix~\ref{sec:appendix} for more discussions. \\
$^b$ Given that TTV singles are statistically similar to transit singles and that TTV multis are statistically similar to transit multis, these two small $p$ values are likely due to a random sampling effect.}
\end{deluxetable*}

To find out how many of these tranets show TTV signals, we cross-match with the TTV catalog of \citet{Holczer:2016}, which was produced based on a uniform search among over 2600 KOIs with relatively high S/N. We find that there are $(23,~12,~7,~4,~0,~1)$ systems in our sample that have $(1,~2,~3,~4,~5,~6)$ tranets and at least one of the tranet shows TTV signals. We dub this the TTV multiplicity function, and also show it in Figure~\ref{fig:multiplicities}.

In our analysis that follows, we will make use of all the transiting planets in our sample, which includes the subsample exhibiting TTV signals. In what follows, we argue that the subsample exhibiting TTVs and those that do not are drawn from a common population, validating our usage of both subsets within a common statistical framework. Based on whether one system has single or multiple tranets and whether any of the tranets show TTV signals, we can divide the whole sample into four subsamples: transit singles, transit multis, TTV singles, and TTV multis. In the transit singles (transit multis) subsample, we do not exclude planets/stars in the TTV singles (TTV multis) subsample. Because the transiting planets outnumber the TTV planets by an order of magnitude, the inclusion or exclusion of the ones with TTVs does not make any noticeable difference. Figure~\ref{fig:comparison} shows the cumulative distributions of planetary (radius and orbital period) and stellar (radius, mass, [Fe/H], and $T_{\rm eff}$) properties of planets/stars in these subsamples, and Table~\ref{tab:p-values} provides the two-sample Kolmogorov-Smirnov (KS) test $p$ values between selected subsamples. There are a number of notable features. First, transit singles and transit multis are statistically similar in almost every index, in particular of the stellar parameters, suggesting that most of the transit singles are likely drawn from the same underlying population and gone through similar formation processes as the transit multis. This is an indication that a substantial fraction of the transit singles are in fact intrinsic multiples, a conclusion that we come to endorse later in the paper.

Another notable feature is that the TTV planets prefer to have slightly larger planetary radii, and are at slightly longer periods, than the rest. This is expected. To enable TTV detections, the planet transits tend to be deeper and have longer periods (\citealt{Lithwick:2012}; see also Equation~(\ref{eqn:ttv-criterion})). However, the stellar properties of systems with and without TTVs (i.e., TTV singles vs. transit singles, and TTV multis vs. transit multis) are statistically similar, since their two-sample KS test $p$ values given in Table~\ref{tab:p-values} are all above the standard threshold (0.05). Therefore, there is no reason to suspect that planets in the TTV sample are drawn from a different population than the tranets are. In this work, we explicitly assume that the TTV planets, despite their relative proximity to mean-motion resonances (MMRs), are not special and that their abundances can be used to constrain the overall planet population. Besides the similarity in stellar parameters, another supporting evidence is that, as new techniques are invented to detect lower-amplitude TTVs, more systems appear to show TTV signals \citep{Ofir:2018}. The transition from showing and not showing TTVs is smooth rather than sharp.

Figure~\ref{fig:multiplicities} shows that, although both the transit and TTV multiplicity functions are monotonically decreasing (subject to statistical noises) as the transit multiplicity increases, the TTV multiplicity function has a weaker dependence on the transit multiplicity. This is because TTV is relatively insensitive to the inclination variations. Although such a feature prevents from using TTV as a characterization technique to precisely constrain the mutual inclination values (e.g., \citealt{Hadden:2017}), it indeed helps to use TTV as a detection technique to probe planet with a broader range of inclination values than the transit technique. The different slopes of the two multiplicity functions are the key to uncover the inclination distribution in multi-planet systems.

\section{Forward Modeling the Observed Multiplicity Functions} \label{sec:model}

The transit and TTV multiplicity functions, as illustrated in Figure~\ref{fig:multiplicities}, are both monotonically decreasing (subject to Poisson noises) as the intrinsic multiplicity $k$ increases, but they behave quantitatively differently, with the TTV multiplicity function less dependent on $k$. Below we show that the transit and TTV multiplicity functions can be simultaneously well described when the inclination dispersion of the $k$-planet system is a power-law function of the intrinsic multiplicity $k$. Using this relation, we can constrain the intrinsic multiplicity vector $\mathbf{F}\equiv (f_1,~f_2,~\cdots,~f_k)$, where $f_k$ is the fraction of Sun-like stars with $k$ \emph{Kepler}-like planets. This section describes the model we use to fit the observed transit and TTV multiplicity functions.

\subsection{Notations} \label{sec:notations}

To facilitate further discussions, we introduce a few mathematical notations here. Following \citet{Tremaine:2012}, we use a matrix $\mathbf{G}$ to quantify the detection probability of planetary systems in transit surveys. Each element $g_{jk}$ denotes the probability that one $k$-planet system is seen to have $j$ ($j\ge1$) tranets. Thus we have
\begin{equation}
    g_{jk} = 0\ ,\quad{\rm if}\ j>k\ .
\end{equation}
If the parameter space that is of interest can fit in at most $K$ planets, then $\mathbf{G}$ should be a $K\times K$ upper-triangular matrix.
\footnote{Note that the \citet{Tremaine:2012} extended their notations to $j=k=0$ and thus their $\mathbf{G}$ matrix had $(K+1)\times (K+1)$ dimensions.}

We use symbol $\mathcal{N}$ for the number of stars in the sample, and the vector $\mathbf{N}=(N_1,~\cdots,~N_K)$ for the observed transit multiplicity function. With the intrinsic multiplicity vector $\mathbf{F}$, the expectation of the transit multiplicity function can be given as
\begin{equation} \label{eqn:transit-expectation}
    \bar{N}_j = \sum_{k=j}^{K} g_{jk} \mathcal{N}f_k = \sum_{k=1}^K g_{jk} \mathcal{N} f_k,\quad {\rm or}~
    \mathbf{\bar{N}} = \mathbf{G} \cdot (\mathcal{N} \mathbf{F})\ .
\end{equation}
The second equality in the summation form has used the fact that $g_{jk}=0$ if $k<j$. The fraction of stars with planets, $F_{\rm p}$, and the average number of planets per star, $\bar{n}_{\rm p}$, are given by
\begin{equation}
    F_{\rm p}=\sum_{k=1}^{K} f_k,\quad \bar{n}_{\rm p}=\sum_{k=1}^K kf_k\ ,
\end{equation}
respectively. The ratio of these two quantities, $\bar{n}_{\rm p}/F_{\rm p}$, gives the average number of planets per planetary system, which we call the \textit{average multiplicity}.

We also introduce the matrix $\mathbf{T}$ to quantify the detection probability of TTVs. Each element $t_{jk}$ represents the probability that one $k$-planet system has $j$ tranets and at least one of them shows detectable TTV signals. The TTV multiplicity function is given by $\mathbf{M}=(M_1,\cdots,~M_K)$, and the expectation of this is given by
\begin{equation} \label{eqn:ttv-expectation}
    \bar{M}_j = \sum_{k=j}^{K} t_{jk} \mathcal{N}f_k,~{\rm or}~
    \mathbf{\bar{M}} = \mathbf{T} \cdot (\mathcal{N} \mathbf{F})\ ;
\end{equation}

\subsection{Model Ingredients} \label{sec:ingredients}

\begin{figure}
\centering
\epsscale{1.2}
\plotone{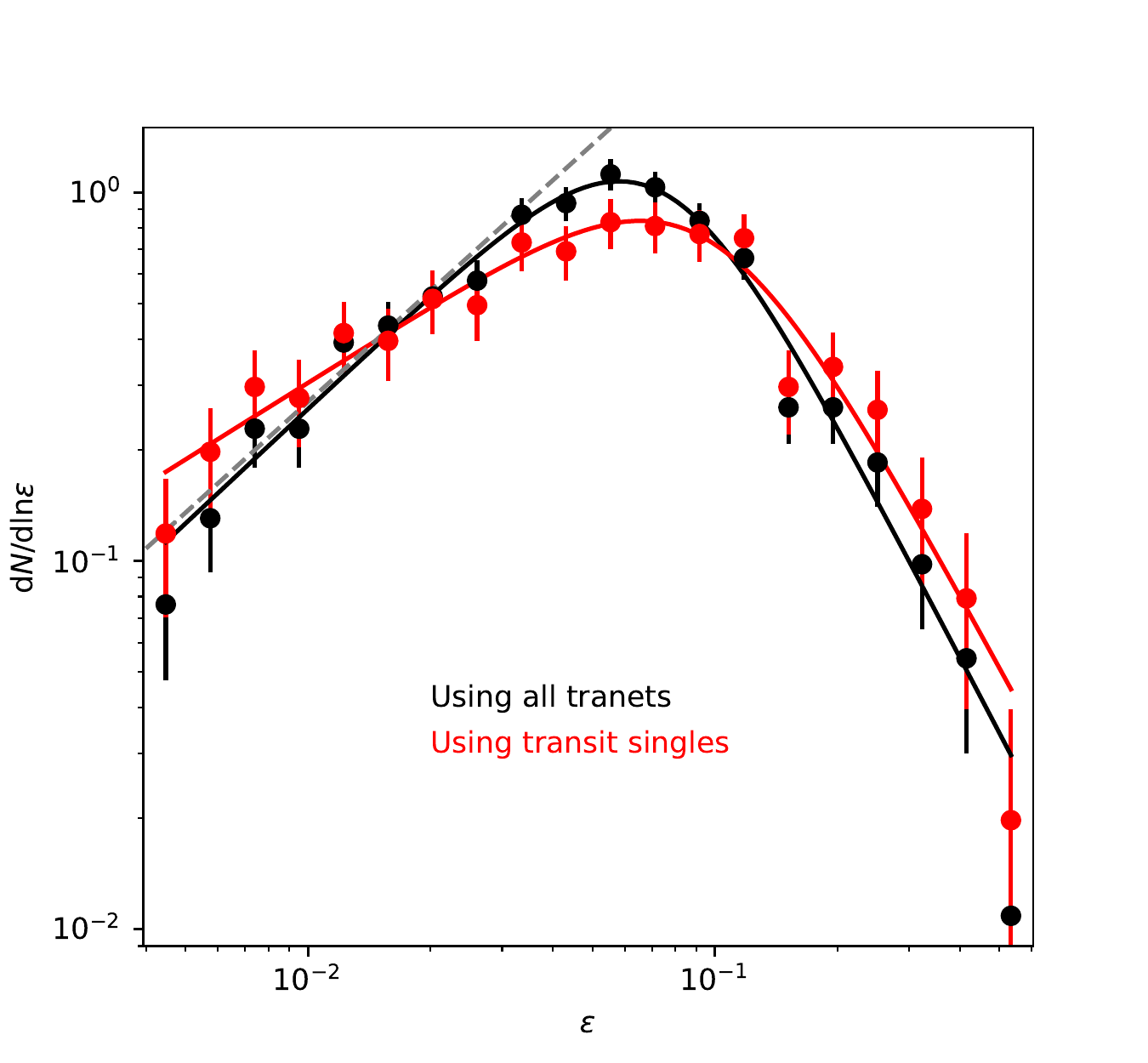}
\caption{Distributions of the transit parameter $\epsilon$ ($\equiv R_\star/a$) using all tranets and only those in transit singles. The gray dashed line marks a logarithmically flat distribution after the correction of the geometric transit probability.
\label{fig:epsilons}}
\end{figure}

In our model, whether a planet transits or not is determined by the transit parameter $\epsilon\equiv R_\star/a$ and its orbital inclination $I_{\rm p}$. We do not take into account the minor impact of the planet size. Below we describe the distributions of $\epsilon$ and $I_{\rm p}$. We also describe the criteria we use in generating multi-planet systems and detecting TTV signals.

\subsubsection{Distribution of Transit Parameters}

Following \citet{Tremaine:2012}, we model the distribution of transit parameter $\epsilon$ as
\begin{equation} \label{eqn:epsilon-general}
    \frac{\dif N}{\dif\ln{\epsilon}} \propto \frac{(\epsilon/\epsilon_0)^a}{1+(\epsilon/\epsilon_0)^b}\ ,
\end{equation}
which is essentially a broken power law but with smooth transition at $\epsilon_0$. Instead of using the values for $\epsilon$ from transit modelings that are not well constrained for some tranets, we re-compute them based on the orbital periods and LAMOST stellar parameters ($R_\star$ and $M_\star$). In this way this parameter $\epsilon$ is better constrained and its lower and upper boundaries are compatible with our sample selection criteria (Section~\ref{sec:sample}). This reconstructed $\epsilon$ distribution using all tranets in our sample is shown as black dots in Figure~\ref{fig:epsilons}, in which we also show the distribution from only transit singles for a reference.

We then model this distribution with the smoothed broken power-law form (Equation~(\ref{eqn:epsilon-general})). After correcting for the geometric transit probability ($\propto \epsilon$), we determine the underlying $\epsilon$ distribution to be
\begin{equation} \label{eqn:epsilon-specific}
    \frac{\dif N}{\dif\ln{\epsilon}} = 0.36 \frac{(\epsilon/\epsilon_0)^{0.04}}{1+(\epsilon/\epsilon_0)^{3.18}} \quad (\epsilon_0=0.074)
\end{equation}
for $0.004<\epsilon<0.6$, and zero elsewhere. This yields a logarithmically flat distribution below $\epsilon_0$, a result in agreement with previous studies \citep[e.g.,][]{DongZhu:2013,Petigura:2013}.

\subsubsection{Distribution of Planetary Inclinations}

For multi-planet systems, the planetary inclination relative to the observer, $I_{\rm p}$, depends on the inclination of the system invariable plane, $I$, the planet inclination with respect to this invariable plane, $i$, and a nuisance parameter $\phi$ (i.e., the phase angle)
\begin{equation} \label{eqn:inc_planet}
    \cos{I_{\rm p}} = \cos{I} \cos{i} - \sin{I}\sin{i}\cos{\phi}\ .
\end{equation}
The distribution of $I$ is isotropic ($\propto \sin{I}$ for $0\le I\le 180^\circ$), and the distribution of $\phi$ is random between $0$ and $360^\circ$. The inclination $i$ quantifies the flatness of the multi-planet system, and we assume that it is related to the number of planets in the system $k$. We parameterize this dependence as a power law between the dispersion of $i$ (or more accurately, $\sin{i}$) and $k$, and choose the normalization at $k=5$
\begin{equation} \label{eqn:delta_inc}
    \sigma_{i,k} \equiv \sqrt{\langle \sin^2 i \rangle} = \sigmai \left(\frac{k}{5}\right)^\alpha\ .
\end{equation}
It is written in this form, so that the normalization factor, $\sigmai$, can be determined separately from the distribution of transit duration ratios of planet pairs in five-planet systems (Section~\ref{sec:normalization}). For such high-multiple planetary systems, the observed multiplicity very likely reflects the intrinsic multiplicity. We decide to use $k=5$ for the normalization term for two reasons. First, there are only a few $k\ge6$ planetary systems found by \emph{Kepler}, the number being so small that the mutual inclination dispersion cannot be well constrained. Second, although there are more four-planet systems than five-planet systems, the fraction of contamination from intrinsically higher-multiplicities is also larger for four-planet systems than for five-planet systems. The power-law index $\alpha$ quantifies the steepness of this inclination dispersion function, and can be constrained from the transit and TTV multiplicity functions.

With this inclination dispersion $\sigma_{i,k}$, the planetary inclination with respect to the invariable plane is then modeled as a \citet{Fisher:1953} distribution \citep{Fabrycky:2009,Tremaine:2012}
\begin{equation} \label{eqn:fisher}
    P(i|\kappa_k) = \frac{\kappa_k \sin{i}}{2\sinh{\kappa_k}} e^{\kappa_k\cos{i}}\ .
\end{equation}
The parameter $\kappa_k$ is related to the dispersion parameter $\sigma_{i,k}$ via
\begin{equation} \label{eqn:kappa}
    \sigma^2_{i,k} = \langle \sin^2{i} \rangle = \frac{2}{\kappa_k} \left(\coth{\kappa_k}-\frac{1}{\kappa_k}\right)\ .
\end{equation}
This Fisher distribution provides a smooth transition from an isotropic distribution ($\kappa_n\ll1$) to a Rayleigh distribution ($\kappa_n\gg1$). The latter one is commonly used for compact multi-planet systems \citep[e.g.,][]{Fabrycky:2014}.

The inclination dispersion $\sigma_{i,k}$, by its definition given by Equation~(\ref{eqn:delta_inc}), has a maximum value of $\sqrt{2/3}$, which can be achieved only when the distribution of $i$ becomes isotropic. For any given $\sigmai$, the upper bound on the inclination dispersion therefore sets a limit on $\alpha$ (and vice versa).

\subsubsection{Stability Criterion} \label{sec:stability}

We describe the stability criterion used in generating multi-planet systems. For intrinsic multiples ($k\ge2$), we inject planets one by one. The transit parameter $\epsilon$ of the first planet is randomly drawn from the distribution specified by Equation~(\ref{eqn:epsilon-specific}), and then the orbital period is derived via
\begin{equation}
    P = \left(\frac{R_\odot}{\rm au}\right)^{3/2} \epsilon^{-3/2} \left(\frac{\rho_\star}{\rho_\odot}\right)^{-1/2} \yr\ .
\end{equation}
Throughout our simulations, we fix $\rho_\star=\rho_\odot$. This is because, the distribution of $\epsilon$ has absorbed the variance of $\rho_\star$, and therefore there is no need to assume a separate distribution for $\rho_\star$. For any additional planet, the transit parameter and the orbital period are randomly assigned in a similar way but with the restriction that the new planet must be far away from any previously injected planets such that the system remains dynamically stable. For the latter, we use the \citet{Deck:2013} stability criterion, which requires that for any given planet pair,
\begin{equation}
    \frac{P_{\rm out}}{P_{\rm in}} > 1+2.2 q^{2/7} = 1.16\ .
\end{equation}
The critical value is derived by assuming a characteristic planet-to-star mass ratio $q=10^{-4}$. The actual choice of this characteristic value has very marginal impact on the modeling output, primarily because of the weak dependence on $q$. Furthermore, even though the chosen $q$ value is larger than the typical planet-to-star mass ratio ($\sim10^{-5}$) of \emph{Kepler} planets, it is extremely rare to have a period ratio below $1.3$ in actual observations \citep{Lissauer:2011,Fabrycky:2014}.

\subsubsection{TTV Detection Criteria} \label{sec:ttv}

For any system with at least one tranet, we determine whether there is detectable TTV signal. Previous studies \citep[e.g.,][]{Holczer:2016,Hadden:2017} have shown that in systems with detected TTVs, it is almost always the case that the TTV signal comes from first-order MMRs between two neighboring planets. We use this empirical result and only consider the closest one (if the tranet is innermost or outermost) or two planets in the TTV detection. We only consider the TTV signals from first-order ($J:J-1=2:1$, $3:2$, $4:3$, and $5:4$) MMRs. Higher-order MMRs are too weak, and the additional first-order MMRs are either not allowed by or too close to the stability limit. We consider the TTV signal to be detectable, if all the following conditions are met for at least one of the chosen first-order MMRs:
\begin{enumerate}
    \item The orbital period of the tranet is less than 200 days.
        \footnote{Although \citet{Holczer:2016} included $P<300~$days in their initial selections, they also required at least six transits observed in the \emph{Kepler} window.}
    \item The super period of the planet pair
        \begin{equation}
            P_{\rm sup}\equiv \frac{P_{\rm in}P_{\rm out}}{|JP_{\rm in}-(J-1)P_{\rm out}|}
        \end{equation}
        is in the range $100-3000$ days. Here $P_{\rm in}$ and $P_{\rm out}$ are the orbital periods of the inner and outer planets, respectively.
    \item The TTV amplitude indicator
        \begin{equation} \label{eqn:ttv-criterion}
            \frac{P}{\Delta} > 1.3\times 10^3~{\rm days}
        \end{equation}
        where $P$ is the orbital period of the tranet, and $\Delta$ is the fractional separation to period commesurability \citep{Lithwick:2012}.
        \begin{equation}
            \Delta \equiv \left| \frac{P_{\rm out}}{P_{\rm in}} \frac{J-1}{J} - 1 \right| \ .
        \end{equation}
\end{enumerate}
The first two criteria are used to mimic the conditions in actual TTV detections \citep{Holczer:2016}. The threshold used in the third criterion, $1.3\times10^3$ days, is approximately the median value of $P/\Delta$ in identified TTV pairs of \citet{Holczer:2016}. For the characteristic mass ratio ($q=10^{-4}$), this corresponds to a 30-min TTV amplitude, which is also the median amplitude of all TTV planets in \citet{Holczer:2016}. Our results are insensitive to the numerical threshold adopted here, as we will show in Section~\ref{sec:discussion}. Specifically, reducing (or increasing) the threshold value on the right-hand side of Equation~(\ref{eqn:ttv-criterion}) will allow more (or less) pairs to become TTV eligible. However, since this raises the fraction of TTV sample uniformly across all systems, this normalization change does not affect our results.

\subsection{Constraining $\sigma_{i,5}$ from Transit Duration Ratios} \label{sec:normalization}

\begin{figure*}
\centering
\epsscale{1.2}
\plotone{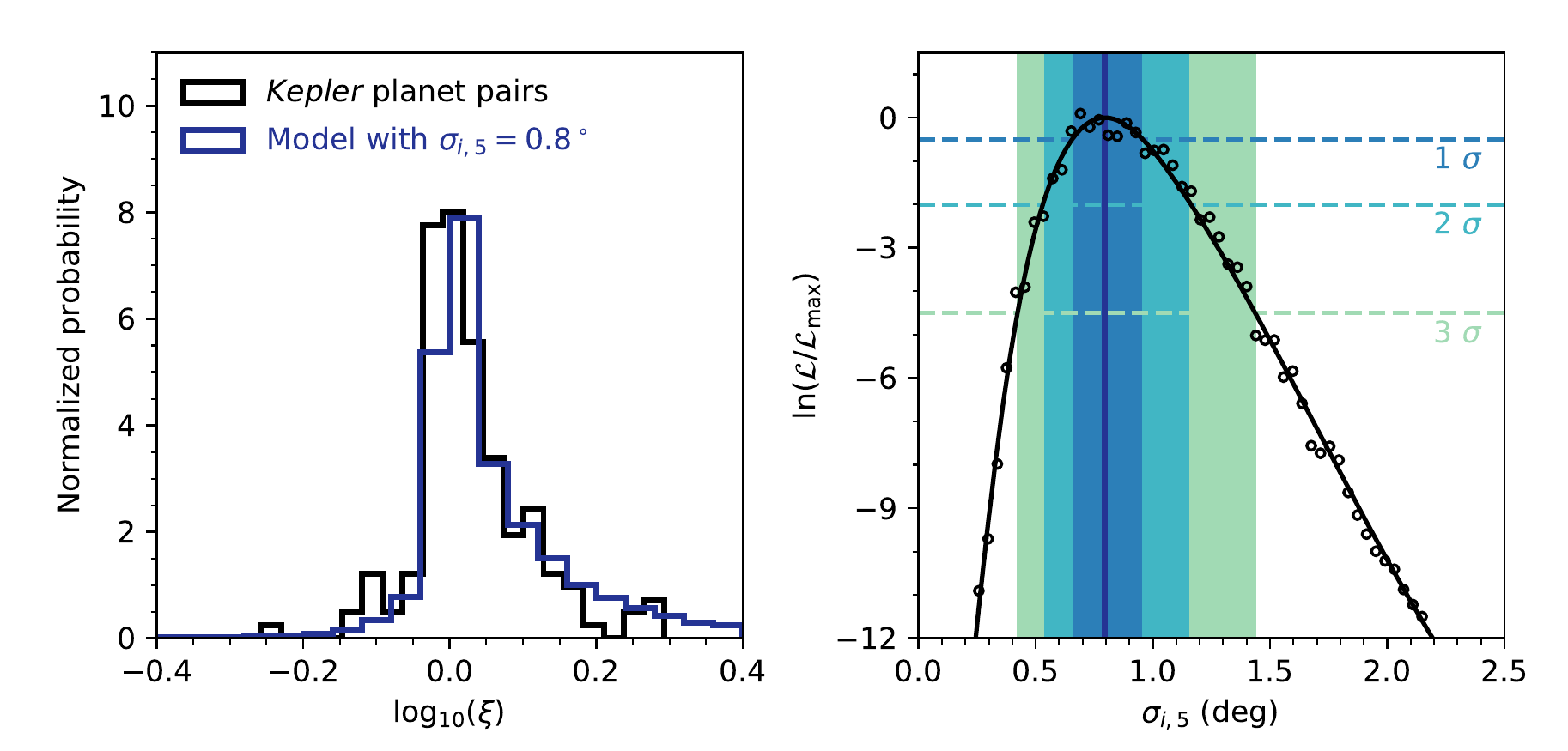}
\caption{Left panel: distributions of the weighted transit duration ratio $\xi$ from a selected sample of 15 five-tranet systems and our best-fit model. Right panel: the likelihood as a function of the inclination dispersion $\sigmai$. The 1-3 $\sigma$ regions are marked out with different colors.}
\label{fig:sig_i5}
\end{figure*}

We acquire external constraints on the normalization factor $\sigmai$ of the inclination dispersion relation (Equation~\ref{eqn:delta_inc}). This is necessary because otherwise we would end up with a strong correlation between $\sigmai$ and $\alpha$.

Because we are constraining $\sigmai$ separately, we are not limited to the planetary systems in our current sample. In fact, our sample only contains three five-tranet systems, and these provide $3\times C_5^2=30$ planet pairs. Instead, we find 15 five-tranet systems whose hosts are Sun-like stars from the California Kepler Survey \citep{Petigura:2017}, which give us 150 planet pairs to constrain the inclination dispersion $\sigmai$.

We adopt a similar approach as \citet{Fabrycky:2014} to constrain the inclination dispersion. Each observed planet pair gives a quantity \citep{Steffen:2010}
\begin{equation} \label{eqn:xi_observable}
    \xi \equiv \frac{T_{\rm dur,in}/P_{\rm in}^{1/3}}{T_{\rm dur,out}/P_{\rm out}^{1/3}}\ ,
\end{equation}
where $T_{\rm dur}$ is the transit duration (from first to fourth contact) and $P$ is the orbital period. The subscripts ``in'' and ``out'' denotes the values of the inner and outer planets, respectively. With Kepler's third law, one can easily see that this parameter $\xi$ is essentially the ratio of orbital-velocity normalized transit durations, which is most sensitive to the inclination $i$ and is marginally dependent on other parameters such as the orbit eccentricity $e$ \citep{Fabrycky:2014}. We can also write $\xi$ in terms of the planet-to-star radius ratio $r$ and the transit impact parameter $b$
\begin{equation} \label{eqn:xi_physical}
    \xi = \sqrt{\frac{(1+r_{\rm in})^2-b_{\rm in}^2}{(1+r_{\rm out})^2-b_{\rm out}^2}}
\end{equation}
Because both $T_{\rm dur}$ and $P$ are much better measured than $r$ or $b$ in observations, the expression given by Equation~(\ref{eqn:xi_observable}) is therefore used in constructing the distribution of $\xi$ from the data. For the 15 five-planet systems, we use the values of $T_{\rm dur}$ and $P$ from the the most recent \emph{Kepler} data release (DR25; \citealt{Thompson:2017}). We then compute the $\xi$ values for individual planet pairs, and show the distribution as the black histogram on the left panel of Figure~\ref{fig:sig_i5}. 

We then model this $\xi$ distribution and attempt to constrain the inclination dispersion $\sigmai$. We ignore the dependence of $\xi$ on the eccentricity $e$ and simply assume circular orbits for all planets in such five-planet systems. While simplifying the modeling, this still remains a very reasonable assumption. First, planets in such high-multiply and stable systems are not expected to have large eccentricities. Second, \citet{Fabrycky:2014} has shown that this $\xi$ parameter alone could not constrain $e$ very well, and that the correlation between inclination dispersion and eccentricity is weak.

For a given value of $\sigmai$, we produce the $\xi$ distribution following the method of \citet{Fabrycky:2014}. First, we randomly draw an impact parameter for the outer planet, $b_{\rm out,sim}$, uniformly from the range $[0,b_{\rm out,max}]$, where $b_{\rm out,max}$ is the impact parameter the planet would require in order that the total S/N ($\sqrt{b_{\rm out,max}/b_{\rm out}}$ times the actual S/N) would be equal to the S/N threshold (7.1). Then we randomly draw $b_{\rm in,sim}$ from a normal distribution with mean $b_{\rm out,sim}(P_{\rm in}/P_{\rm out})^{2/3}$ and dispersion $\sigmai/(d_{\rm in}+r_{\rm in})$, where $d\equiv R_\star/a$ is the stellar radius scaled to the planet-star separation and is taken as the value from DR25. Such a normal distribution in impact parameters reproduces a Rayleigh distribution with dispersion $\sigmai$ in inclinations. If $|b_{\rm in,sim}|\le b_{\rm in,max}$, we consider this simulated planet to be detectable. Otherwise we repeat the previous step until the above condition is met. Once a simulated planet pair is generated, we compute $\xi$ using Equation~(\ref{eqn:xi_physical}). We repeat the whole process and generate 250 simulated pairs for each planet pair in the sample. The $\xi$ distribution for the given $\sigmai$ is then generated from the ensemble of all simulated pairs.

We then compute the likelihood for the actual $\xi$ distribution to be drawn from the simulated one. We note that this is more accurate than the approach of \citet{Fabrycky:2014}, which used the $p$-value of Kolmogorov-Smirnov test. We compute the likelihood as
\begin{equation}
    \mathcal{L} = \prod_{j=1}^{N} \int P_{\rm sim}(\ln{\xi}) \exp{\left[-\frac{(\ln{\xi_j}-\ln{\xi})^2}{2\sigma_{\ln\xi,j}^2}\right]} \mathrm{d}\ln\xi\ ,
\end{equation}
where $P_{\rm sim}(\ln{\xi})$ is the probability distribution of simulated $\xi$ in logarithmic space, $\xi_j$ is the observed value of $\xi$ of the $j$-th planet pair, and $\sigma_{\ln\xi,j}$ is the fractional uncertainty of $\xi_i$. We compute $\sigma_{\ln\xi,j}$ based on the measured uncertainties on $T_{\rm dur,in}$ and $T_{\rm dur,out}$.

We repeat the Monte Carlo simulation and compute the likelihood $\mathcal{L}$ for 100 values of $\sigmai$ equally spaced from $0.1^\circ$ to $4.0^\circ$, and show the results on the right panel of Figure~\ref{fig:sig_i5}. As the scatter plot shows, the likelihood reaches its maximum around $\sigmai=0.8^\circ$. The simulated $\xi$ distribution for this best-fit $\sigmai$ is also shown on the left panel of Figure~\ref{fig:sig_i5}. To find the different confidence levels of $\sigmai$, we use a spline function to smooth the $\ln{(\mathcal{L}/\mathcal{L}_{\rm max})}$ vs. $\sigmai$ scatter plot, and find that the 1-$\sigma$ 2-$\sigma$, and 3-$\sigma$ confidence intervals, defined as $\ln(\mathcal{L}/\mathcal{L}_{\rm max}) \geq -n^2/2$ to be $0.65^\circ-0.96^\circ$, $0.53^\circ-1.16^\circ$, and $0.42^\circ-1.44^\circ$ for $n=1$, 2, and 3, respectively.

For the subsequent modeling of the power-law index $\alpha$ and intrinsic multiplicity vector $\mathbf{F}$, we will only consider values of $\sigmai$ from the 3-$\sigma$ confidence interval. The smoothed $\ln{(\mathcal{L}/\mathcal{L}_{\rm max})}$ is also used as our prior on $\sigmai$ unless specified otherwise.

\subsection{Monte Carlo Simulations}

We use Monte Carlo simulations to compute the $\mathbf{G}$ and $\mathbf{T}$ matrices for a grid of $\sigmai$ and $\alpha$. We note that \citet{Tremaine:2012} provided an analytical formalism to compute the $\mathbf{G}$ matrix. However, the time-limiting factor is always the computation of the $\mathbf{T}$ matrix, which may not be done in the analytical way. We compute the $\mathbf{T}$ matrix within a Monte Carlo, and this automatically produces the $\mathbf{G}$ matrix.

For the intrinsic singles ($k=1$), there is no TTV signal ($t_{11}=0$), and only one parameter ($g_{11}$) needs to be calculated. This can be done analytically \citep{Tremaine:2012}: $g_{11}=\langle \epsilon \rangle = 0.03$.

For intrinsic multiples ($k\ge2$), we first randomly draw the inclination of the invariable plane, $I$, from an isotropic distribution, and then inject planets one by one with the stability criterion given in Section~\ref{sec:stability} imposed. For each planet, we assign randomly a phase parameter, $\phi$, from the uniform distribution and the planet inclination with respect to the invariable plane, $i$, from the Fisher distribution (Equation~(\ref{eqn:fisher})), whose parameter $\kappa_k$ is determined by Equations~(\ref{eqn:delta_inc}) and (\ref{eqn:kappa}). The inclination of the planet with respect to the line of sight is then given by Equation~(\ref{eqn:inc_planet}). Finally, whether a planet is a tranet or not is determined by the transit criterion
\begin{equation}
    \epsilon>\cos{I_{\rm p}}\ .
\end{equation}
For any system with at least one tranet, we invoke the TTV criteria in Section~\ref{sec:ttv} to determine whether there is detectable TTV signals.

With the above procedures, we are able to generate an $k$-planet system, compute the number of tranets, and determine whether any of the tranets shows detectable TTV signals. Given that we only have up to six tranets in one system, our simulations run up to six-planet systems. The impact of higher multiples to our results will be discussed in Section~\ref{sec:fraction}. We repeat the whole process and generate a large number of planetary systems, until each element in the $\mathbf{T}$ matrix is determined to $<2\%$ precision.

\subsection{Modeling the Observed Multiplicity Functions}

For a given intrinsic multiplicity vector $\mathbf{F}$ and matrices $\mathbf{G}$ and $\mathbf{T}$, the probability to see the transit and TTV multiplicity functions as shown in Figure~\ref{fig:multiplicities} is given by
\begin{equation}
    \mathcal{L} = \prod_{k=0}^K \frac{\bar{N}_k^{N_k} \exp(-\bar{N}_k)}{N_k!} \times \prod_{k=1}^K \frac{\bar{M}_k^{M_k} \exp(-\bar{M}_k)}{M_k!}\ .\,
\end{equation}
where $\bar{N}_k$ and $\bar{M}_k$ are given by Equations~(\ref{eqn:transit-expectation}) and (\ref{eqn:ttv-expectation}), respectively. Note that here we have extended the transit multiplicity function down to $k=0$, with $f_0=1-\sum_{k\ge1} f_k$. In practice, we ignore the constants in the log of the above likelihood, and try to maximize the following quantity to find the best model parameters ($\sigmai$, $\alpha$, and $\mathbf{F}$)
\begin{equation}
    \ln{\mathcal{L}} = \sum_{k=0}^K (N_k \ln{\bar{N}_k} - \bar{N}_k) + \sum_{k=1}^K (M_k \ln{\bar{M}_k} - \bar{M}_k)\ .
\end{equation}
Given the large number of dimensions, we choose the Markov Chain Monte Carlo (MCMC) algorithm that is implemented in \texttt{emcee} \citep{emcee:2013} as the optimization method. This way we also obtain the posterior distributions of $F_k$ ($k=1,~\cdots,~6$) for given sets of $\sigmai$ and $\alpha$.

\section{Results} \label{sec:results}

\begin{figure*}
\centering
\epsscale{1.2}
\plotone{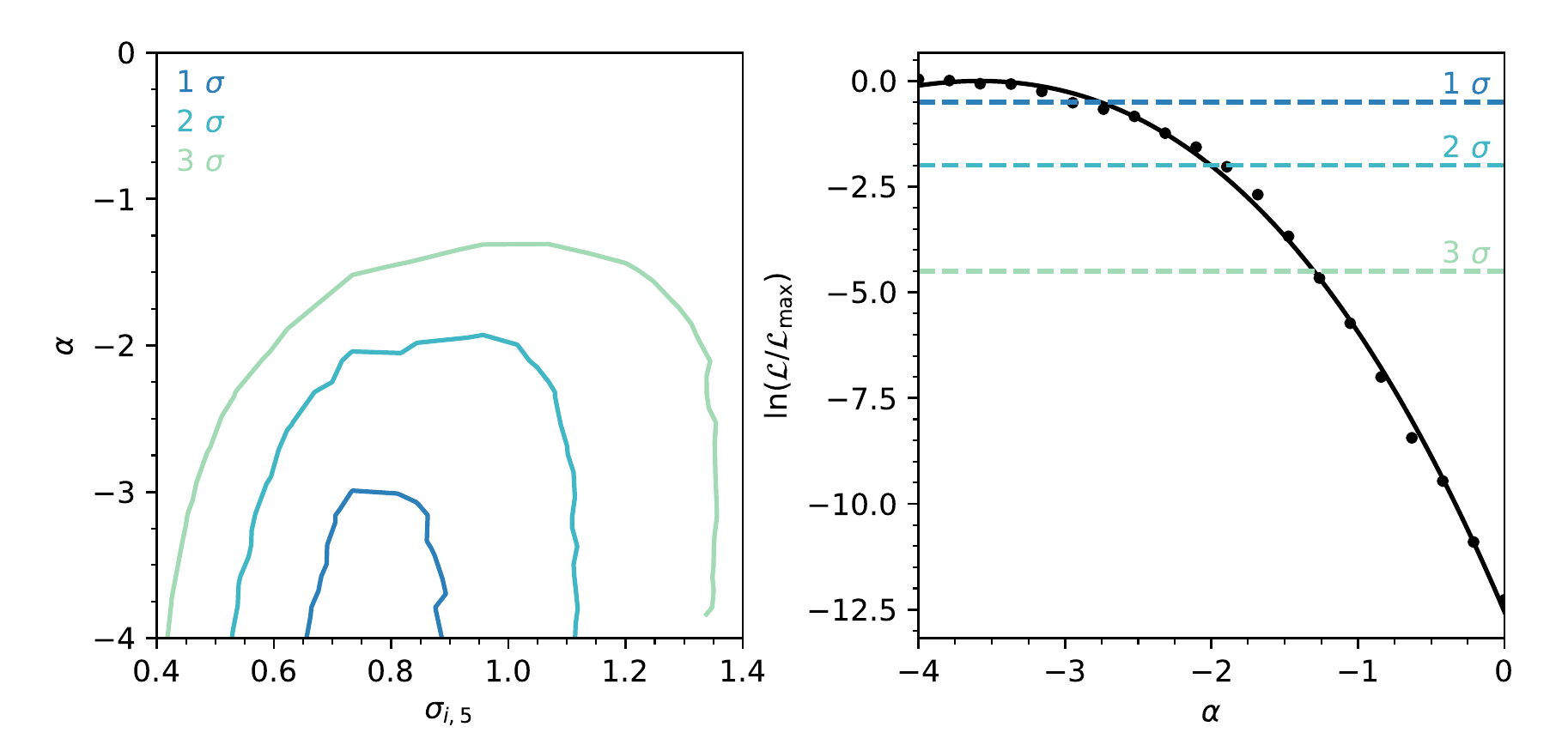}
\caption{Posterior distributions of the parameters quantifying the inclination dispersion function. The left panel shows the 1-3 $\sigma$ contours in the power-law index $\alpha$ vs. normalization factor $\sigmai$ plane. The 3-$\sigma$ contour does not fully reach $\alpha=-4$ line at the rightmost end, because the lower limit of $\alpha$ at $\sigmai=1.35^\circ$ is slightly above $-4$. The right panel shows the $\ln{\mathcal{L}}$ as a function of the power-law index $\alpha$. The black dots are the calculations from grid points, and the black curve is the smoothed function. Here 1-3 $\sigma$ regions are also indicated.
\label{fig:posteriors}}
\end{figure*}

\begin{figure}
\centering
\epsscale{1.2}
\plotone{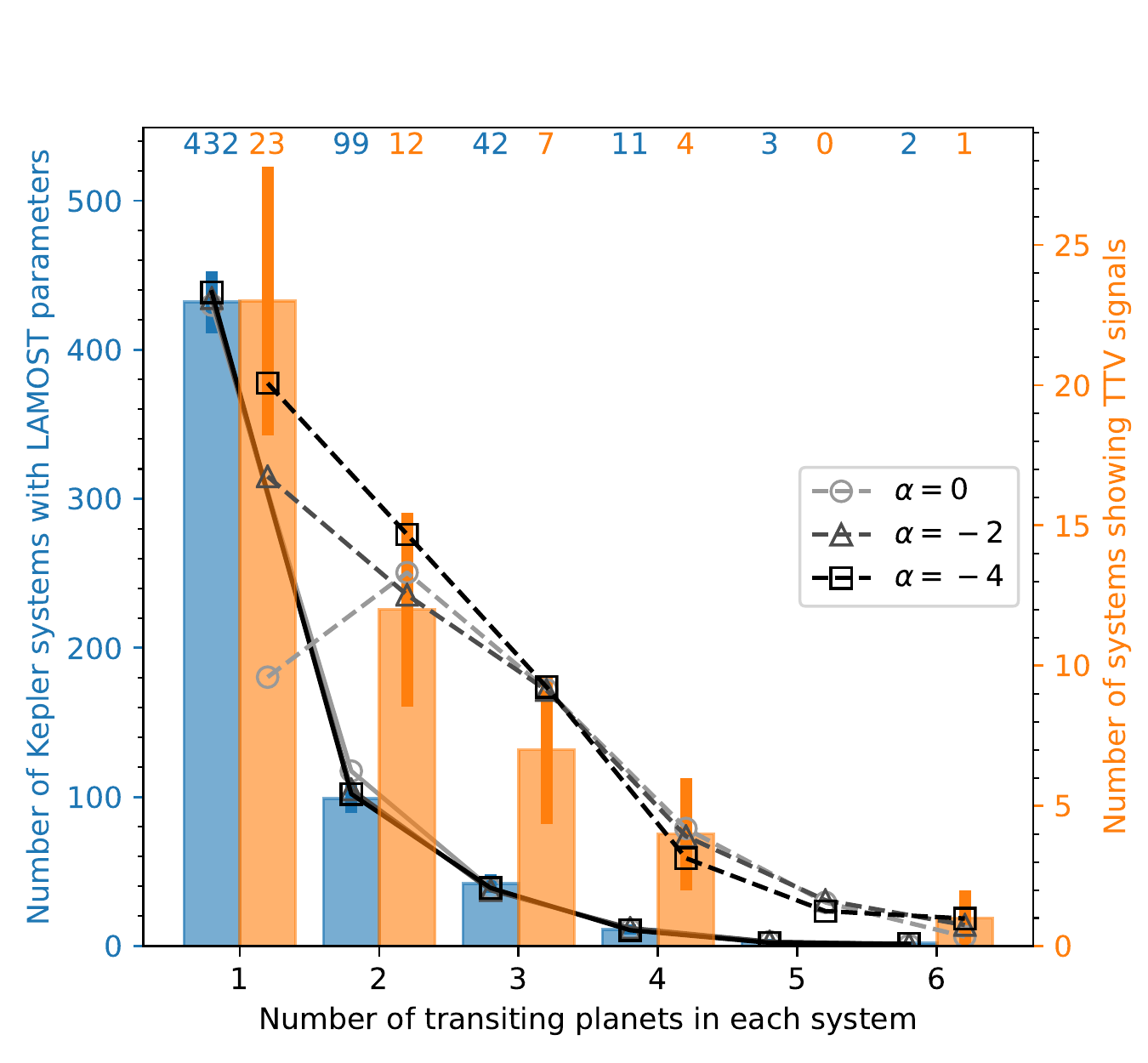}
\caption{The observed multiplicity functions (with error bars) and the best-fit models with different values of the power-law index $\alpha$. The numbers on the top indicate the actual numbers in individual bins. We use solid lines for models of the transit multiplicity function and dashed lines for models of the TTV multiplicity function.
\label{fig:best-fit}}
\end{figure}

\begin{figure*}
\centering
\epsscale{1.2}
\plotone{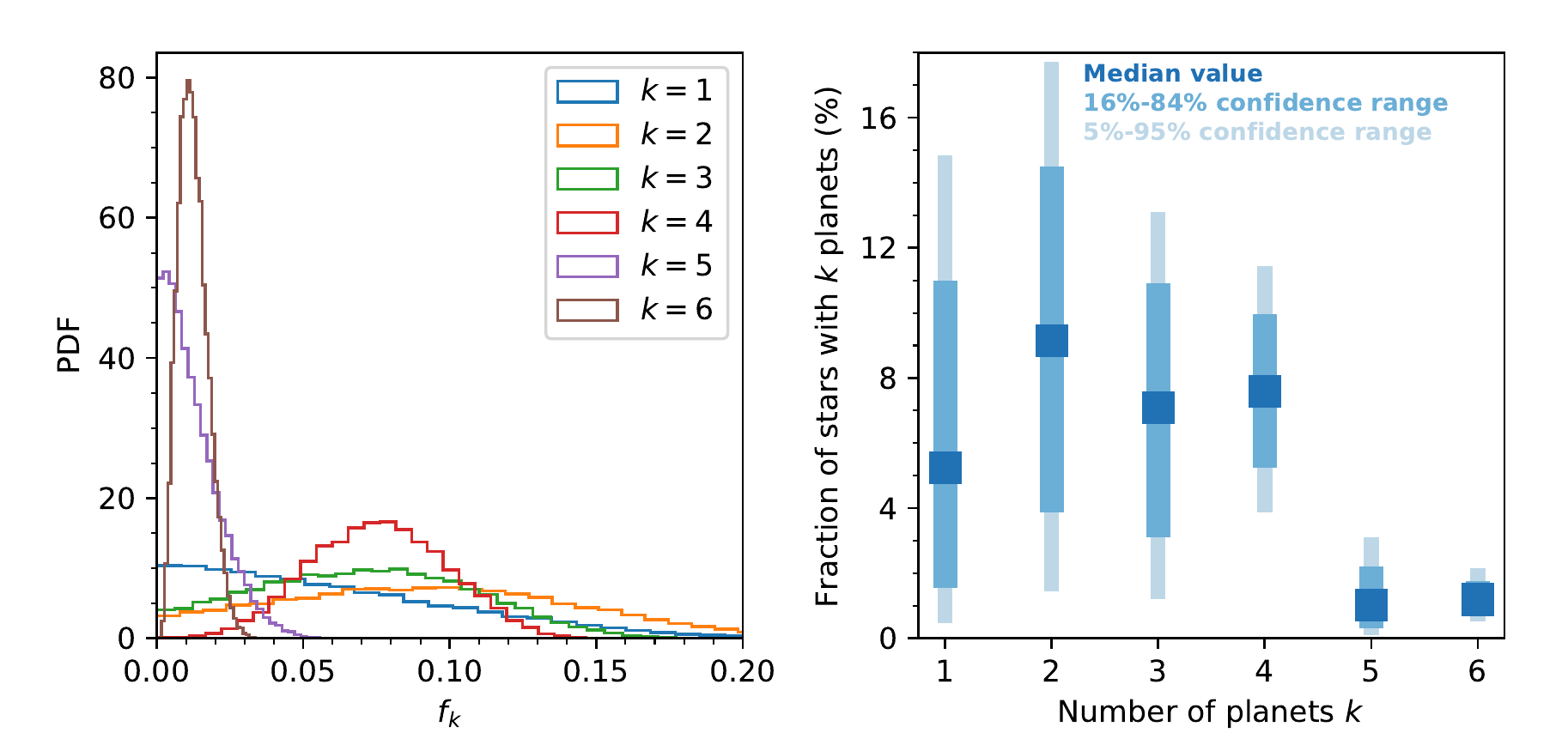}
\caption{Constraints on the individual components $f_k$ of the intrinsic multiplicity vector $\mathbf{F}$. The left panel shows the full posterior distributions, and the right panel shows the reported values and associated uncertainties.
\label{fig:intrinsic}}
\end{figure*}

\begin{figure*}
\centering
\epsscale{1.2}
\plotone{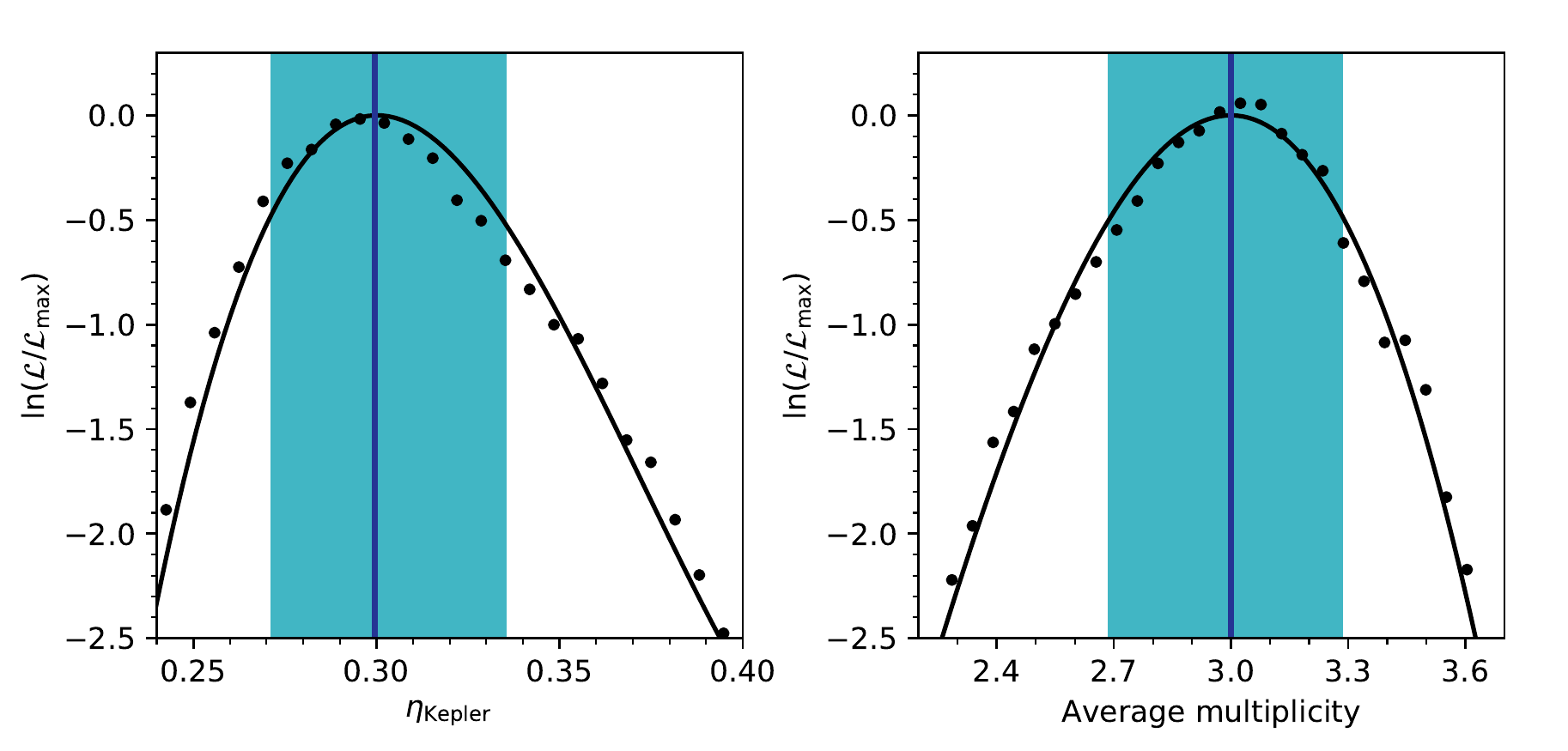}
    \caption{Constraints on the fraction of Sun-like stars with \emph{Kepler}-like planets (left panel) and the average number of planets per \emph{Kepler} planetary system (right panel). The black dots are the (normalized) $\ln{\mathcal{L}}$ values from the model, and the black curve is the smoothed distribution. The vertical line and vertical bands indicate the best values and 1-$\sigma$ uncertainties.
\label{fig:overall-occurrence}}
\end{figure*}

\subsection{Planetary Inclination Dispersion}

We find the maximum likelihood values for each grid point in the ($\sigmai,~\alpha$) plane following the procedure detailed in the previous section. After imposing the prior probability on $\sigmai$ from the transit duration ratios (Figure~\ref{fig:sig_i5}), we can determine the $n$-$\sigma$ contours, where $n$ ($=1,~2,~3$) refers to the number of $\sigma$ and the contour is defined by $\ln(\mathcal{L}/\mathcal{L}_{\rm max})=-n^2/2$. Consequently, we can construct the posterior distribution of the power-law index $\alpha$ in a similar way. These results are shown in Figure~\ref{fig:posteriors}.

Figure~\ref{fig:posteriors} indicates that the power-law index $\alpha$ is well constrained to be close to its lower bound $-4$, which is given by the normalization factor $\sigmai\approx0.8^\circ$ and $\sigma_{i,2}\le\sqrt{2/3}$ (Section~\ref{sec:ingredients}). Specifically, $\alpha<-3$ at 1-$\sigma$ level, $\alpha<-2$ at 2-$\sigma$ level, and $\alpha\ge0$ can be securely excluded. Therefore, the more planets a system has, the smaller the planetary inclination dispersion is, and this dispersion is a steep function of the intrinsic multiplicity.

To understand what information is driving the constraint on $\alpha$, we show in Figure~\ref{fig:best-fit} the best-fit models with fixed values of $\alpha$. As this figure shows, a steep inclination dispersion function is required primarily because of the large number of TTV singles, which contribute nearly half of systems with TTVs. Although our sample only contains 23 TTV singles, the ratio between numbers of TTV singles and TTV multiples remains essentially the same even if a much larger TTV catalog \citep{Holczer:2016} or a different TTV catalog \citep{Ofir:2018} is used. Therefore, our constraints on the power-law index parameter is robust.

Figure~\ref{fig:best-fit} seems to suggest that there may be more TTV singles than even the steepest model curve ($\alpha=-4$) can account for. However, with our current sample size it is not statistically significant that would require a more complicated model than our current one.

\subsection{Intrinsic Multiplicity Vector}

We stack the Markov chains from all MCMC runs, and discard entries that are more than 3-$\sigma$ away ($\Delta \chi^2>9$) from the best one. With this combined Markov chain, we can then investigate the constraints on individual components of the intrinsic multiplicity vector $\mathbf{F}$.
\footnote{We confirm that this approach produces very similar but smoother posteriors than directly using the shape of the $\chi^2$ (i.e., $\ln{\mathcal{L}}$) curve.}

The left panel of Figure~\ref{fig:intrinsic} shows the full posterior distributions of individual components $f_k$, which is the fraction of Sun-like stars with $k$ ($1\le k \le 6$) \emph{Kepler}-like planets. As we can see, the majority of the components are not constrained very well, because of the strong degeneracies between neighboring components. However, it is notable that the first component, $f_1$, is consistent with zero. That is, there can be effectively zero intrinsic singles, and nearly all the transit singles are in fact intrinsic multiples with the additional planets non-transiting. This is a result we have discussed qualitatively in Section~\ref{sec:introduction}.

For future practical usage (such as to predict the yield of future transit missions), we nevertheless report measurements and uncertainties of individual components $f_k$. This is done by taking the median, $16\%-84\%$, and $5\%-95\%$ of the posterior distributions. The result is shown in the right panel of Figure~\ref{fig:intrinsic}. Note that this plot seems to suggest a sharp drop in occurrence rate from low multiples ($k\le4$) to high multiples ($k\ge5$), but this feature is artificial and comes purely from the way these values are derived.

\subsection{Overall Planet Occurrence Rates} \label{sec:overall-occurrence}

Although the individual components of the intrinsic multiplicity vector $\mathbf{F}$ are not well constrained, the overall occurrence rates, meaning the total fraction of stars with planets $F_{\rm p}$ and the average number of planets per star $\bar{n}_{\rm p}$, are found to be well constrained. This is not surprising for $\bar{n}_{\rm p}$, because this quantity only depends on the distribution of the transit parameter $\epsilon$ \citep{Youdin:2011,Tremaine:2012}. In fact, as we prove in Appendix~\ref{sec:appendix}, 
\begin{equation} \label{eqn:number-indicator}
    \bar{n}_{\rm p}=\frac{\sum_{j=1}^K jN_j}{\mathcal{N} \langle \epsilon \rangle}\ .
\end{equation}
That is, the average number of planets per star is given by the total number of tranets, the total number of stars, and the average probability that one planet transits. For our sample, the above equation gives $\bar{n}_{\rm p}=0.90\pm0.03$, which agrees with our model outputs. Our constraint on $\bar{n}_{\rm p}$ also agrees with previous studies \citep[e.g.,][]{Fressin:2013,Petigura:2013}.

As the left panel of Figure~\ref{fig:overall-occurrence} indicates, the total fraction of Sun-like stars with \emph{Kepler}-like planets is also well constrained. To avoid the confusion with the general fraction $F_{\rm p}$ (for arbitrary planet sizes and orbital distances), we introduce $\eta_{\rm Kepler}$ for this specific fraction.
\footnote{This notation follows the well-accepted term $\eta_\oplus$.}
The posterior distribution gives
\begin{equation}
    \eta_{\rm Kepler} = 30\pm3\%\ .
\end{equation}
This is a factor of $\sim$two lower than previous estimates \citep{Fressin:2013,Petigura:2013,Winn:2015}. With the determinations of $\bar{n}_{\rm p}$ and $\eta_{\rm Kepler}$, we also find that on average each \emph{Kepler}-like planetary system has $3.0\pm0.3$ planets, as shown in the right panel of Figure~\ref{fig:overall-occurrence}. Our work is the first to determine this average multiplicity.

How could the total fraction be constrained so well even though the individual components $f_k$ were not? We will provide the explanation in Section~\ref{sec:fraction}, but the conclusion is that, this results from some property of the matrix $\mathbf{G}$ that relates the nature of the transit probabilities. Therefore, our result that only $30\%$ of Sun-like stars host \emph{Kepler}-like planets is robust, and in particular, it is not sensitive to the details of TTV multiplicity function or our TTV modelings.

\section{Discussion} \label{sec:discussion}

\subsection{The Determination of $\eta_{\rm Kepler}$} \label{sec:fraction}

\begin{figure*}
\centering
\epsscale{1.2}
\plotone{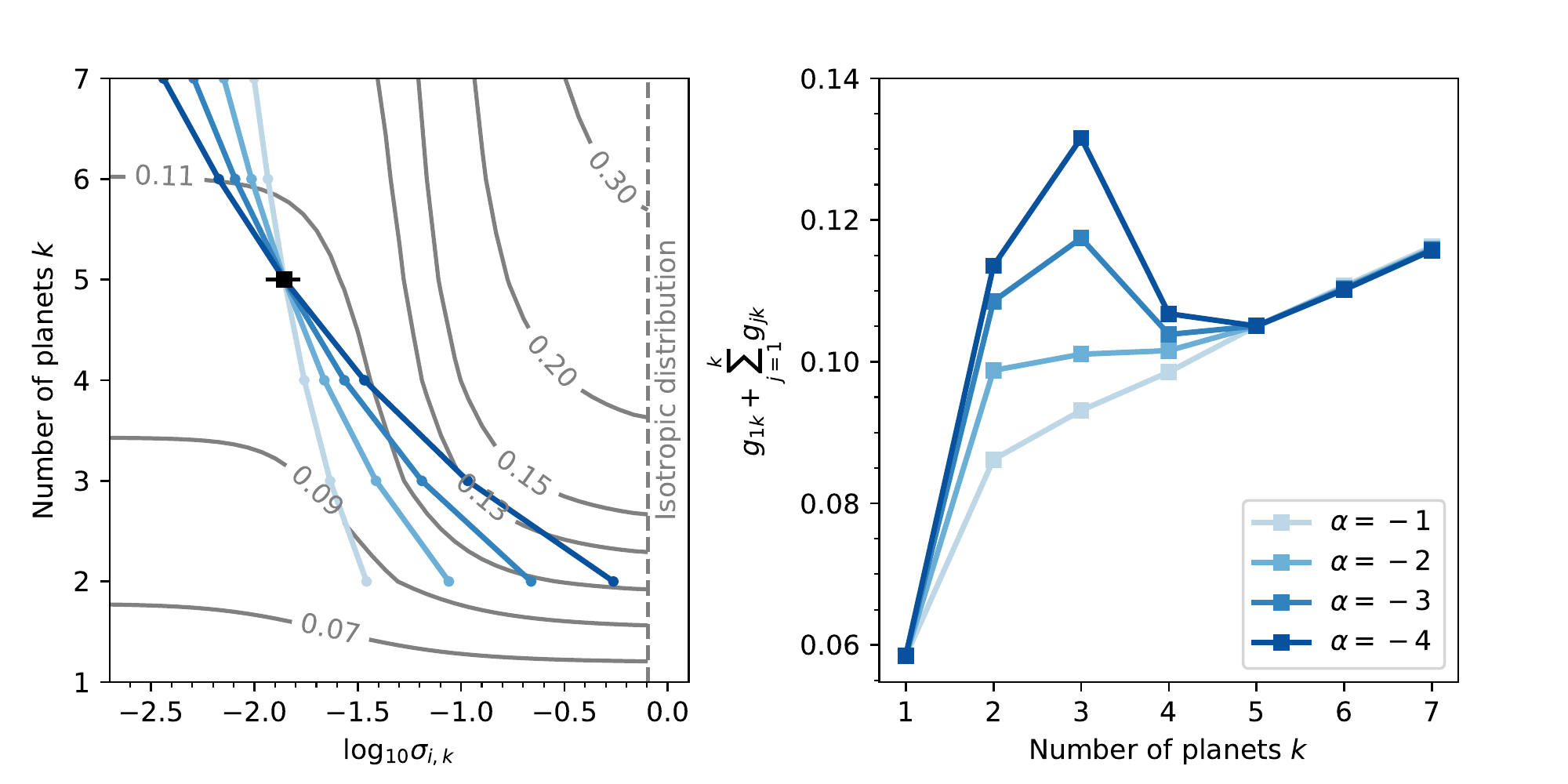}
\caption{Left panel: the gray contours show the values of the quantity, $g_{1j}+\sum_{j=1}^k g_{jk}$, for different combinations of $k$ and $\sigma_{i,k}$. The black square with error bar is our determination of the normalization factor $\sigmai$. The blue-ish curves are the inclination dispersion functions with different values of $\alpha$. Right panel: different curves are the quantity $g_{1j}+\sum_{j=1}^k g_{jk}$ as a function of $k$ for different values of power-law index $\alpha$.
\label{fig:gmat}}
\end{figure*}

Here we answer the question from Section~\ref{sec:overall-occurrence}: why $\eta_{\rm Kepler}$ could be constrained so well?

Following Section~\ref{sec:notations}, the total number of transiting systems reads
\begin{equation}
    \sum_{j=1}^K \bar{N}_j = \mathcal{N} \sum_{k=1}^K f_k \sum_{j=1}^k g_{jk}\ .
\end{equation}
Unfortunately, the quantity $\sum_{j=1}^k g_{jk}$ (i.e., the probability to see at least one tranet) is not a constant. Otherwise the determination of $F_{\rm p}$ would be straightforward. However, unless the period distribution of planets in multiple systems is dramatically different from the period distribution of planets in single systems, the probability to have at least one tranet out of $k$ ($k\ge 1$) planets is no less than the probability to see one tranet in the single-planet systems. Mathematically, this implies $\sum_{j=1}^k g_{jk} \ge g_{11}$. Therefore, the above equation gives an upper limit on $F_{\rm p}$
\begin{equation}
    F_{\rm p} \equiv \sum_{k=1}^K f_k \le \frac{1}{\mathcal{N} g_{11}} \sum_{k=1}^K \bar{N}_k\ .
\end{equation}
For our sample, this gives $\eta_{\rm Kepler} \le 62\%$.

A second relevant quantity is the number of transit singles
\begin{equation}
    \bar{N}_1 = \mathcal{N}\sum_{k=1}^K f_k g_{1k}\ .
\end{equation}
Again, the quantity $g_{1k}$, the probability to see one tranet in a $k$-planet system, is not conserved: the more planets a system has and the hotter (i.e., larger inclination dispersion) the system is, the larger this quantity will be.

However, we find that the combination of the two quantities, $g_{1k}+\sum_{j=1}^k g_{jk}$, remains roughly constant for broad ranges of $\sigmai$, $\alpha$, and $k$. 
This can be seen in Figure~\ref{fig:gmat}, which illustrates the dependence of this quantity on various model parameters. Here because only the $\mathbf{G}$ matrix is involved, we use the deterministic approach of \citet{Tremaine:2012} to compute $\mathbf{G}$. This approach requires to truncate the Legendre series at a certain threshold $l_{\rm max}$. Unlike \citet{Tremaine:2012} who used $l_{\rm max}=50$, we choose a much higher threshold, $l_{\rm max}=3000$, which is necessary in order to have the elements of $\mathbf{G}$ for flat and multi-planet systems (i.e., upper left corner of the left panel of Figure~\ref{fig:gmat}) converge.

The conservation of the quantity $g_{1k}+\sum_{j=1}^k g_{jk}$ means that, the more planets a system has and the hotter the system is, the larger this joint probability is, and the two separate probabilities, $g_{1k}$ and $\sum_{j=1}^k g_{jk}$, compensate each other if the number of planets increases while the inclination dispersion decreases. Figure~\ref{fig:gmat} also suggests that the weighted mean
\begin{equation}
    \langle g_{1k}+\sum_{j=1}^k g_{jk} \rangle \equiv \frac{1}{F_{\rm p}} \sum_{k=1}^K f_k \left( g_{1k}+\sum_{j=1}^k g_{jk} \right) \approx 0.11\ ,
\end{equation}
so that the total fraction of stars with planets can be given by
\begin{equation} \label{eqn:fraction-estimator}
    F_{\rm p} = \frac{1}{\mathcal{N} \langle g_{1k}+\sum_{j=1}^k g_{jk} \rangle} \left( N_1 + \sum_{j=1}^K N_j \right)\ ,
\end{equation}
With numbers from our sample, this estimator directly gives $\eta_{\rm Kepler}=30\%$, as long as the following three conditions are met.

First, the $k=1$ term does not provide significant contribution to the weighted mean. That is, there is no large population of intrinsic singles. This has been discussed qualitatively in Section~\ref{sec:introduction}, and here we provide a simple quantitative argument. Of all tranets in our sample, 432 and 395 are in transit singles and transit multis, respectively. Of the subset of those that show detected TTV signals, 23 are in transit singles and 26 (excluding the double counting from tranet pairs both showing TTVs) are in transit multis. For the transiting planets in transit multis, the probability of showing TTV is $26/395=6.6\%$. For those in transit singles, the same probability is $23/432=5.3\%<6.6\%$, which we interpret as the blending of intrinsic singles in the transit singles. One therefore finds that at most 19\% (or 82 systems) of the transit singles are intrinsic singles. With the mean transit probability ($\langle \epsilon \rangle=0.03$) and the total number of surveyed stars (30759), the fraction of intrinsic singles ($f_1$) is at most $9\%$. Although this upper limit is derived based on a subset of the \emph{Kepler} catalog and a specific TTV table \citep{Holczer:2016}, the conclusion remains the same even if one uses the multiplicity fraction of the overall \emph{Kepler} catalog or a different TTV table \citep{Ofir:2018}. Even with this upper limit for $f_1$, the weighted mean of $g_{1k}+\sum_{j=1}^K g_{jk}$ only varies by $10\%$.

Second, the more planets there are in the system, the smaller the planet inclination dispersion is. Together with previous investigations of the multi-tranet systems \citep{Lissauer:2012,Tremaine:2012,Fabrycky:2014}, the result that there is no large population of intrinsic singles suggests that a significant fraction of planetary systems must have large planet-planet mutual inclinations. Although it is in principle possible that these are high multiples, it is much more likely that such systems with high mutual inclinations are low multiples, and indirect evidence from the planet eccentricity study also supports this \citep{Xie:2016}.

Finally, there is no large population of very high ($k\ge7$) multiples. Given the previous result, these $k\ge7$ multiples should be very flat. With our inner ($\sim1$ day) and outer (400 days) period boundaries, the weighted quantity $f_{1k}+\sum_{j=1}^K f_{jk} \lesssim 0.25$, and the probability to see seven tranets is at least $R_\odot/(1.06~{\rm au})=0.0044$. Therefore, the fact that we do not have any 7-tranet system in a sample of 30,759 stars sets an upper limit on the fraction of systems with $k\ge7$, $f_{\ge7} < 2.2\%$ (95\% confidence level). Including these very high multiples in the calculation of the weighted mean changes it by $10\%$ in the opposite direction as the intrinsic singles does.

The above arguments explain why the total fraction can be constrained very well even though the individual components cannot, and more importantly, confirm that our determination of the total fraction of Sun-like stars with \emph{Kepler}-like planets, $\eta_{\rm Kepler}$, is fairly robust, and does not depend on the details of our modeling or the TTV multiplicity function. Furthermore, because the average number of planets per star, $\bar{n}_{\rm p}$, is determined independently from the inclination distribution (Equation~(\ref{eqn:number-indicator})), the average multiplicity is also robustly measured.

Equation~(\ref{eqn:fraction-estimator}) also points toward a robust and straightforward way to determine the total fraction of stars with planets, which has practical applications. As \citet{Zhu:2016} have pointed out, the fraction of stars with planets is better than the average number of planets per star for the purpose of quantifying the correlation between planet formation efficiencies and stellar properties (such as metallicity). It is nevertheless the latter that has been broadly used.

\subsection{Only 30\% of Sun-like Stars Host \emph{Kepler} Planets}

In Section~\ref{sec:introduction} we have explained qualitatively why previous studies \citep{Fressin:2013,Petigura:2013} overestimated $\eta_{\rm Kepler}$. Now we explain it in a more quantitative way. We focus on the transit studies here, but the conclusion should apply to RV studies \citep[e.g.,][]{Mayor:2011} as well given that the same statistical approach was used.

There are two primary differences in our study and previous studies: the parameter space under investigation and the statistical method. Here we discuss the impact of the former. We use \citet{Fressin:2013} as the example of previous studies. \citet{Fressin:2013} took into account both the geometric transit probability and the pipeline detection efficiency, and concluded that 52\% of Sun-like stars should have at least one planet with $R_{\rm p}>0.8~R_\oplus$ and $P<85$ days. Our result that only 30\% of Sun-like stars host \emph{Kepler} planets comes by studying the \emph{Kepler} planets as a whole and only accounting for the geometric transit probability. However, the parameter space we study is inclusive of the parameter space investigated in \citet{Fressin:2013}. Specifically, the pipeline detection efficiency would only increase the number of systems with planets of $R_{\rm p}>0.8~R_\oplus$ and $P<85$ days by 10\%. Such a small change is far from what is needed to explain the discrepancy between our work and \citet{Fressin:2013}.

All previous studies on the fraction of stars with planets used the probability to detect the most detectable planet for the probability to detect at least one planet. In those transit studies \citep{Fressin:2013,Petigura:2013}, by reducing the number of tranets in all systems to one, they assumed that the resulting average number of planets per star should be the total fraction of stars with planets. Using the distribution of the transit parameter $\epsilon$ determined by transiting planets in transit singles (the red curve in Figure~\ref{fig:epsilons}),
\footnote{Because transit singles outnumber the transit multiples significantly. The $\epsilon$ distribution will be essentially the same regardless of whether the innermost tranets \citep{Fressin:2013} or the most detectable tranets \citep{Petigura:2013} of transit multiples are included.}
we find that the probability that a typical planet transits is $g_{11}=0.025$. According to Equation~(\ref{eqn:number-indicator}) the resulting average number of planets per star is 0.77. Taking this value for $\eta_{\rm Kepler}$ would mean an overestimation by a factor of $2.6$. Note that this value (77\%) also exceeds the upper limit (62\%) we derived in Section~\ref{sec:fraction} under the very general condition.

\begin{figure}
\centering
\epsscale{1.2}
\plotone{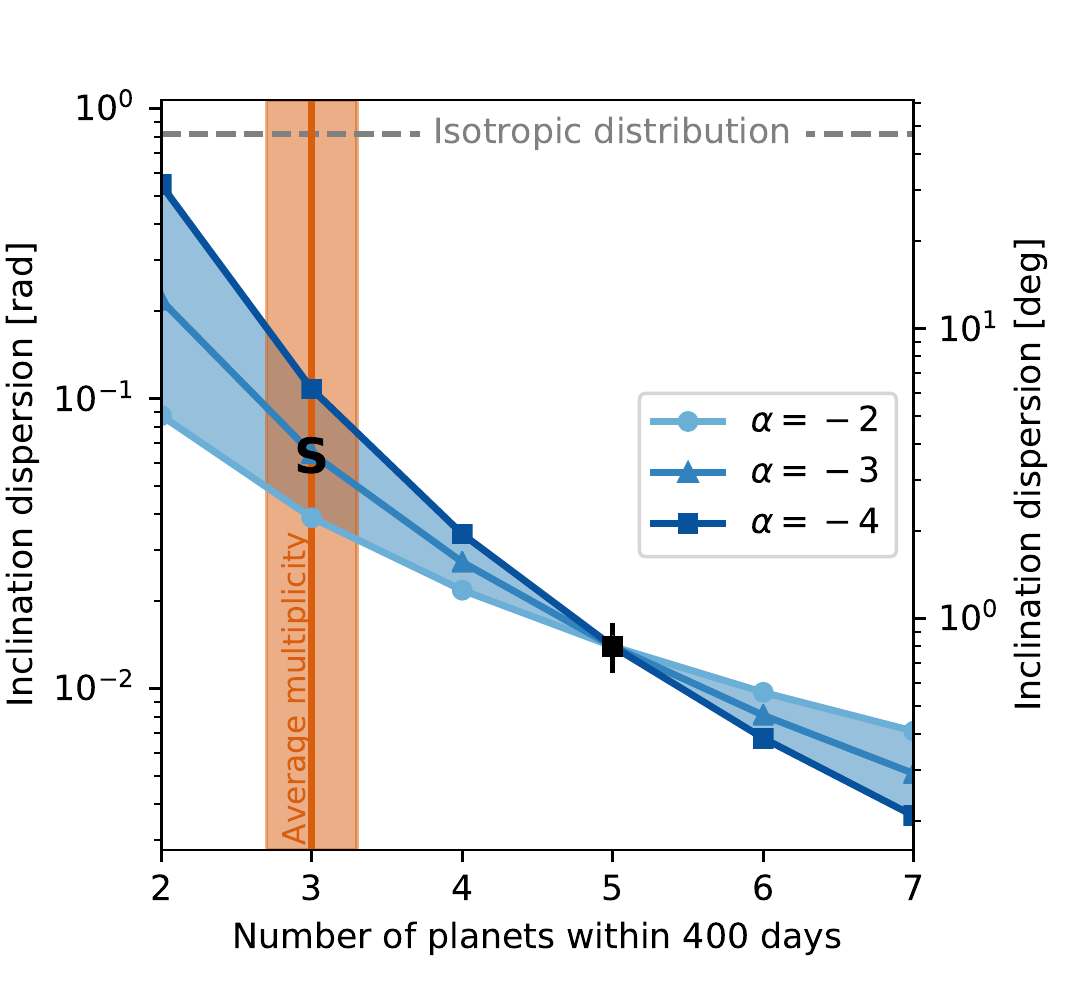}
\caption{The blue-ish curves show the inclination dispersion as a function of intrinsic multiplicity, with slightly different colors for different values of power-law index $\alpha$. Here we only show the 2-$\sigma$ range $-4\le \alpha \le-2$. The black square with error bar indicates the normalization factor we obtain from the transit duration ratios of five-tranet systems. The orange vertical band shows the 1-$\sigma$ range of the average multiplicity, with the central vertical line indicating the best value. The gray dashed line is the upper limit of inclination dispersion, which can be achieved if the inclination follows an isotropic distribution. We also mark the position of our Solar system with `S' in this plot. Throughout our simulations, we always use the radian values of the inclination dispersion, and they are also easy to be connected to the orbital eccentricities \citep{Xie:2016}. However, for discussions of the inclination itself, especially at small values, we also indicate the scales in degree.
\label{fig:summary}}
\end{figure}

\subsection{Inclination Dispersions and Multiplicities: Comparisons with Theories}

The bulk of the \emph{Kepler} planets are the so-called super-Earths (planets with radii between Earth and Neptune), and the majority of such planets discovered by \emph{Kepler} reside within $\sim$100 days. Such super-Earths are absent in our own Solar system. We find that the average number of such planets a system hosts is $\sim 3$, and that the dispersion of planetary inclinations is a steep function of the intrinsic multiplicity. As Figure~\ref{fig:summary} shows, a system with $k\ge5$ planets within 400 days is pretty flat, with inclination dispersion within $1^\circ$. But as the number of planets reduces, the system puffs up and the planets can be mutually inclined up to $\sim$10$^\circ$. An isotropic distribution for planets in two-planet systems is not completely ruled out by the data. Together with the statistical result of eccentricities from transit durations \citep{Xie:2016}, we now have a consistent picture that systems with fewer planets are dynamically hotter. In the following, we briefly cast these results in light of formation theories that have been proposed.

The formation of super-Earths remain unresolved. In  the standard core accretion theories \citep{IdaLin:2004,Mordasini:2009}, in which the cores of these planets are formed at large distances when the gas disk is still fully present, these planets are migrated rapidly to near the inner edge of the proto-planetary disks. It is thought that MMRs naturally get set up between planets that are migrating with different speeds, preventing them from being engulfed by the host stars. However, such a story predicts an abundance of MMRs, in contrast with the observed period ratios in which MMRs feature minimally \citep{Lissauer:2011,Fabrycky:2014}. While multiple scenarios have been proposed to break the planets out of MMRs \citep{GS2015,Delisle2014,CF2015,Liu2017}, it is unclear that the overall period distribution can be explained. It is also unclear, in such a framework, how to account for the diverse multiple systems and their current dynamical states (but see \citealt{Izidoro:2017}).

An alternative scenario, first proposed by \citet{Hansen2013}, favors {\it in-situ} formation, in which these planets are locally assembled. 
\footnote{Here, migration by the gas disk is artificially suppressed. One possibility for this is late assembly \citep{lee2014}.}
This is similar to the conventional story for the formation of terrestrial planets (e.g., \citealt{kokubo1998,kokubo2000,raymond2004}). Using $N$-body simulations, \citet{Hansen2013} studied planet assembly, starting from a large number of proto-planets with a total mass of $20~M_\oplus$ within $400$ days. They further assumed that all impacts are fully accretional, and that there is no external source of dissipation (by gas or by planetesimals). They presented a number of statistical properties for the resulting systems, which we proceed to compare against here.  

First, \citet{Hansen2013} found that the median number of planets in a system is $4$, an average eccentricity dispersion of $\sigma_e = 0.11$, and an average inclination dispersion $\sigma_i \sim$ a few degrees. These are somewhat similar to our findings of $3$ planets per system, and $\langle kf_k \sigma_{i,k} \rangle \sim 4^\circ$ ($\langle f_k \sigma_{i,k} \rangle \sim 6^\circ$) with $2\leq k\leq7$. Second, \citet{Hansen2013} noted that their simulations predicted too few single tranet systems. This remains a problem even after this work. The mismatch seems to be mainly due to the differences in their predicted multiplicities: $f_1=f_2=0$ and $f_3\ll f_4,f_5$ in \citet{Hansen2013}, compared to $f_k\sim5-10\%$ for $k=1,2,3,4$ and much smaller for $k>4$ in our work (Figure 8).

We show that lower-$k$ systems are dynamically hotter \citep[also see][ for the eccentricity aspect]{Xie:2014}. What is the reason behind this? 

One possibility is that this is a natural outcome of the stability requirement. One naively expects that a more packed system (higher $k$) has to have a lower dispersion to avoid dynamical instability. From \citet{Pu:2015} the critical number of mutual Hill radii that a system can have is $K_{\rm crit}=2+k+27/5\times \sigma_{i,k}[3/(2q)]^{1/3}$, with $q=M_p/M_\star$. Here we have assumed that $\sigma_e\sim2\sigma_i$ and that the dependence of $K_{\rm crit}$ on $\sigma_{i,k}$ obtained for $k=5$ can be applied to smaller $k$ values. 
\footnote{The extrapolation might break down at $k=2$ as $K_{\rm crit}$ is somewhat insensitive to changes in mutual inclinations for $\lesssim40^\circ$ \citep{petrovich2015}.}
Further assuming that the systems extend one decade ($\sim 0.1-1$ AU) in semi-major axis, then we find that the total number of Hill spheres is $\ln(10)[3/(2q)]^{1/3}$, and that the average separation (in unit of Hill radii) between planets in a $k-$planet system is $\ln(10)[3/(2q)]^{1/3}/k$. By equating this average separation and the critical number $K_{\rm crit}$, we get that the critical inclination dispersion is $\sigma_{i, k} \sim  24^\circ/k-0.8^\circ(1+k/2)$ (with $q=10^{-5}$). This boundary follows a decreasing trend with $k$, which is similar to but less steep than our result in Figure~\ref{fig:summary}. However, given the approximate nature of the arguments above, we are unable to quantitatively compare this prediction with our result, specifically at $k=2$ where the extrapolation from larger $k$ likely breaks down. Further theoretical work will be required to address whether stability is responsible for the observed inclination dispersion trend.

It is also possible that the correlation between $\sigma_{i,k}$ and $k$ reflects the formation process via giant impacts. Unfortunately, \citet{Hansen2013} did not explicitly present their eccentricity/inclination dispersions as functions of intrinsic multiplicities, which would have made a pivotal comparison against our Figure~\ref{fig:summary}. An analytic version of this formation process was put forward by \citet{tremaine2015}, which used the ergodic approximation and dynamical stability. This model predicts that the dispersion in eccentricities $\propto1/k$, which is similar to the stability argument given above if we assume $\sigma_i\propto\sigma_e$. In practice, this model does not provide any prediction for the inclinations as it has assumed coplanar orbits.

Some theoretical works suggest that distant giant planets could drive up the mutual inclinations of the inner planetary systems and/or decrease their multiplicity by dynamical instabilities \citep{Johansen:2012,Lai:2017,Huang:2017,Mustill:2017,Pu:2018}. If this is the primary channel for the observed features, one would expect that the stellar hosts of single tranets (low $j$ and/or large $\sigma_i$) should reflect the well-known giant planet-metallicity correlation \citep[e.g.,][]{Fischer:2005}. However, as the lower middle panel of Figure~\ref{fig:comparison} shows, the metallicity distribution of the stars with single tranets is similar to that of stars with multiple tranets. A similar result is found by looking at the sample of single-tranet systems with TTVs, which should be indicative of multi-planet systems with large inclination dispersions. Therefore, these observations indicate that the perturbation by the unseen distant giant planet is likely not the primary cause of the observed inclination dispersions.

Alternatively, host stars with a large quadrupole moment can also excite the planet-planet mutual inclinations \citep{Spalding:2016,Spalding:2018}. Unfortunately our current sample does not have enough hot ($T_{\rm eff}>6200$ K) stars, for which this mechanism is expected to be most efficient, to test this hypothesis.

\subsection{Solar System \textit{vs.} \emph{Kepler} Systems}

We have determined the intrinsic architecture of \emph{Kepler} planetary systems, and now we discuss briefly how our Solar system fits in this revised picture.

First of all, with $\eta_{\rm Kepler}=30\%$, our Solar system belongs to the majority of Sun-like stars that do not have any planet detectable by \emph{Kepler}. As was first pointed out by \citet{Chiang:2013}, the \emph{Kepler} systems contain more solid mass in their inner regions than is expected from a minimum-mass Solar nebula (MMSN). The typical mass for \emph{Kepler} planets is $\sim3~M_\oplus$ \citep{Marcy:2014,Fulton:2017,Hadden:2017,Owen:2017}. So these systems contain from 3 to 18 $M_\oplus$ of solid masses, in contrast to our own $2~M_\oplus$ (within 400 days). As far as the inner ($\lesssim 1$ au) planetary system is concerned, it seems plausible that the difference arises because our system is a slightly low-metallicity version of the typical \emph{Kepler} system.
\footnote{The Solar system has giant planets beyond 1 au, which is atypical for Sun-like stars with Solar metallicity. If the outer planetary system is also taken into account, our Solar system no longer belongs to the majorities \citep{Zhu:2018}.}

Even though none of the Solar system planets is detectable by \emph{Kepler}, it is interesting to notice that our own system shares at least two similarities with the \emph{Kepler} systems, as has been shown in Figure~\ref{fig:summary}. First, similar to the typical \emph{Kepler} system, we also have three planets within 400 days. Second, if only considering these three planets (Mercury, Venus, and Earth), the average inclination relative to the invariable plane, $3.5^\circ$, is consistent with the inclination dispersions of 3-planet \emph{Kepler} systems. It is important to note, however, that the higher-mass Kepler systems are typically thought to form in a substantially shorter amount of time (prior to the dispersal of the natal disk) as compared to the Solar system terrestrial planets \citep{Morbidelli:2012,Chiang:2013}. It is therefore unclear whether the similarities between the Solar System and Kepler systems are purely coincidental or are representative of a more fundamental behavior of general planetary systems.


\acknowledgements
We thank the anonymous referee for comments. We would also like to thank Andy Gould for comments on an earlier version of the manuscript.
This paper includes data collected by the \emph{Kepler} mission. Funding for the \emph{Kepler} mission is provided by the NASA Science Mission directorate.
This paper also uses data from the LAMOST survey. Guoshoujing Telescope (the Large Sky Area Multi-Object Fiber Spectroscopic Telescope, LAMOST) is a National Major Scientific Project built by the Chinese Academy of Sciences. Funding for the project has been provided by the National Development and Reform Commission. LAMOST is operated and managed by the National Astronomical Observatories, Chinese Academy of Sciences.
CP acknowledges support from the Jeffrey L. Bishop Fellowship and from the Gruber Foundation Fellowship. 
SD acknowledges Project 11573003 supported by National Natural Science Foundation of China (NSFC).

\appendix
\section{Simple Estimation of Average Number of Planets Per Star} \label{sec:appendix}

\begin{figure}
\centering
\epsscale{1.2}
\plotone{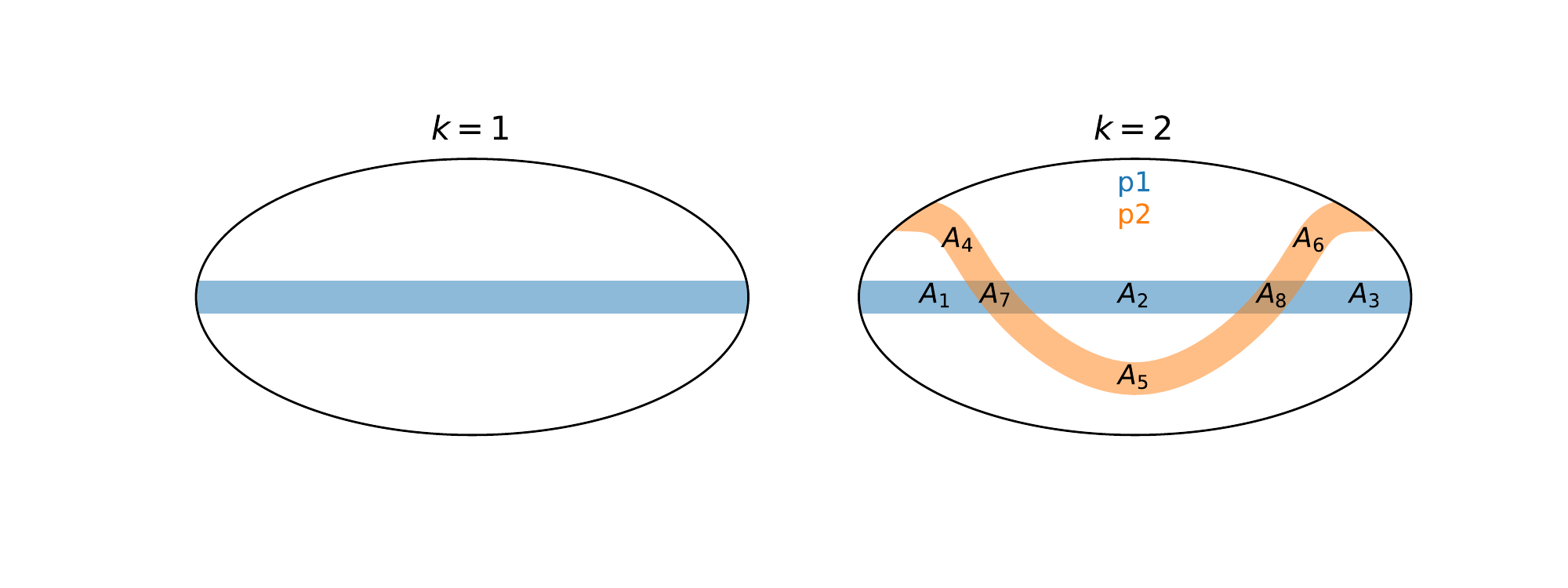}
    \caption{Schematic views of the transit probabilities for 1-planet (left panel) and 2-planet (right panel) cases. The unit sphere is now projected onto a plane following the standard Mollweide projection. The shaded regions indicate the positions of the observer where the transit happens. The width of the band corresponds to the transit parameter $\epsilon$. In the 2-planet case, the overlapping regions are the positions where the observer would see both planets transit. We use symbols $A_i$ ($i=1,~\cdots,~8$) to denote the area of individual strips.
\label{fig:schematic}}
\end{figure}

As \citet{Youdin:2011} pointed out, the average number of planets per star $\bar{n}_{\rm p}$ is directly given by the the total number of tranets $\sum_j j N_j$, the total number of stars $\mathcal{N}$, and the average transit parameter $\langle\epsilon \rangle$. See also our Equation~(\ref{eqn:number-indicator}). With our mathematical notations, this is equivalent to
\begin{equation} \label{eqn:equality}
    \sum_{j=1}^k jg_{jk} = k\langle \epsilon \rangle \ .
\end{equation}
In other words, the number of tranets one expects to see in a $k$-planet system is proportional to $k$, regardless of the details of $g_{jk}$. This can be proved mathematically using the expression of $\mathbf{G}$ in \citet{Tremaine:2012}. Below we provide another simple and robust proof, and discuss the associated assumption.

The transit probability is essentially the fractional area on a unit sphere where transit happens ($\epsilon>\cos{I_p}$). This can be shown graphically by projecting the 3D sphere onto a 2D plane, as done in Figure~\ref{fig:schematic} for 1-planet and 2-planet cases. See also \citet{Brakensiek:2016} for the 3-planet case. For $k=1$, Equation~(\ref{eqn:equality}) reduces to the definition of $\langle \epsilon \rangle$. For $k=2$, with the notations for different areas in Figure~\ref{fig:schematic} the left-hand-side of Equation~(\ref{eqn:equality}) is $(A_1+A_2+A_3+A_4+A_5+A_6)+2(A_7+A_8)=(A_1+A_7+A_2+A_8+A_3)+(A_4+A_7+A_5+A_8+A_6)=g_{11}(p1) + g_{11}(p2)$, where $p1$ and $p2$ denote the planet 1 and 2, respectively. Therefore, the left-hand-side equals to the right-hand-side as long as the transit parameters (or approximately the separations) of the two planets are statistically no different. This is the only assumption that goes into Equation~(\ref{eqn:equality}). The $k\ge3$ cases can be easily proved similarly.

Is the separation distribution of planets in singles different from the separation distribution of planets in multiples? If using tranets, one indeed sees a difference between these two distributions. For example, see the upper middle panel of Figure~\ref{fig:comparison} as well as Figure~\ref{fig:epsilons}. However, this is not super surprising, as the detections of tranets around the same star are correlated due to the geometric effect. Even so, the difference in these two distributions is not very prominent. Perhaps a better sample to use is the sample of planets from RV observations. As \citet{Tremaine:2012} have discussed, the distributions of the semi-major axes of RV planets are statistically the same in single- and multiple-planet systems. See in particular the lower left panel of their Figure~2.


\end{CJK*}
\end{document}